\documentclass[10pt,journal,commsoc]{IEEEtran}
\ifCLASSINFOpdf
  \usepackage[pdftex]{graphicx}
\else
\fi

\usepackage{courier}
\usepackage{amsmath}
\usepackage{amssymb}
\usepackage{amsfonts}
\usepackage{xcolor}
\usepackage{multirow}
\usepackage{float}
\usepackage{graphicx}
\usepackage{graphics}
\usepackage{subfigure}
\usepackage{float}
\usepackage{bigstrut}
\usepackage{booktabs}
\usepackage{bibentry}
\usepackage{balance}
\usepackage{amsmath}

\begin{document}

\title{ICE: Information Credibility Evaluation on Social Media via Representation Learning}

\author{Qiang~Liu,
Shu~Wu,~\IEEEmembership{Member,~IEEE,}
Feng~Yu,
Liang~Wang,~\IEEEmembership{Senior Member,~IEEE,}
and~Tieniu~Tan,~\IEEEmembership{Fellow,~IEEE}
\IEEEcompsocitemizethanks{\IEEEcompsocthanksitem The authors are with the Center for Research on Intelligent Perception and Computing (CRIPAC), National Laboratory of Pattern Recognition (NLPR), Institute of Automation, Chinese Academy of Sciences (CASIA) and the University of Chinese Academy of Sciences (UCAS), Beijing, 100190, China.
\protect\\
E-mail: \{qiang.liu, shu.wu, feng.yu, wangliang, tnt\}@nlpr.ia.ac.cn.}

\thanks{}}
\markboth{Journal of \LaTeX\ Class Files,~Vol.~14, No.~8, August~2015}%
{Shell \MakeLowercase{\textit{et al.}}: Bare Demo of IEEEtran.cls for Computer Society Journals}

\IEEEtitleabstractindextext{%
\begin{abstract}
With the rapid growth of social media, rumors are also spreading widely on social media, such  as microblog, and bring negative effects to human life. Nowadays, information credibility evaluation has drawn attention from academic and industrial communities. Current methods mainly focus on feature engineering and achieve some success. However, feature engineering based methods often require a lot of labor and cannot fully reveal the underlying relations among data. In our viewpoint, the key elements of evaluating credibility are concluded as \textit{\textbf{who}}, \textit{\textbf{what}}, \textit{\textbf{when}}, and \textit{\textbf{how}}. These existing methods cannot well model the correlation among these key elements during the spreading of microblogs. In this paper, we propose a novel representation learning method, Information Credibility Evaluation (ICE), to learn representations of information credibility on social media. In ICE, latent representations are learnt for modeling \textit{who}, \textit{what}, \textit{when}, and \textit{how}, and these key elements means user credibility, behavior types, temporal properties, and comment attitudes respectively. The aggregation of these factors in the microblog spreading process yields the representation of a user's behavior, and the aggregation of these dynamic representations generates the credibility representation of an event spreading on social media. Besides, in ICE, a pairwise learning method is applied to maximize the credibility difference between rumors and non-rumors. To evaluate ICE, we conduct a series of experiments on a Sina Weibo dataset, and the experimental results show that the proposed ICE model outperforms the state-of-the-art methods.
\end{abstract}

}

\maketitle

\IEEEdisplaynontitleabstractindextext

%
\IEEEpeerreviewmaketitle

\section{Introduction}
\IEEEPARstart{W}{ith} the rapid growth of social media, such as Facebook, Twitter, and Sina Weibo, people are sharing information and expressing their attitudes publicly. Social media brings great convenience to users, and information can be spread rapidly and widely nowadays. However, rumors can also be spread on the Internet more easily. A rumor is an unverified and instrumentally relevant statement of information spreading among people \cite{difonzo2007rumor}. Rumors bring significant harm to daily life, social harmony, or even public security. With the growth of the Internet and social media, such harm will also grow greater. For instance, as the loss of MH370 has drawn worldwide attention, a great amount of rumors has spread on social media, e.g., MH370 has landed in China,\footnote{http://www.fireandreamitchell.com/2014/03/07/rumor-malaysia-airlines-mh370-landed-china/} the loss of MH370 is caused by terrorists,\footnote{http://www.csmonitor.com/World/Asia-Pacific/2014/0310/Malaysia-Airlines-flight-MH370-China-plays-down-terrorism-theories-video} and Russian jets are related to the loss of MH370.\footnote{http://www.inquisitr.com/1689765/malaysia-airlines-flight-mh370-russian-jets-in-baltic-may-hold-clue-to-how-flight-370-vanished/} These rumors about MH370 mislead public attitudes to a wrong direction and delay the search of MH370. Up to October 10, 2015, on the biggest Chinese microblog website Sina Weibo,\footnote{http://weibo.com} 28,454 rumors have been reported and collected in its misinformation management center.\footnote{http://service.account.weibo.com/?type=5\&status=4} Accordingly, it is crucial to evaluate information credibility and to detect rumors on social media.

To automatically evaluate information credibility on social media, some methods have been recently proposed. Existing methods are mainly based on feature engineering, i.e., methods with handcrafted features. Most of them are based on content information and the source credibility at the microblog level \cite{castillo2011information}\cite{qazvinian2011rumor}\cite{gupta2013faking} or event (containing a group of microblogs) level \cite{kwon2013prominent}\cite{zhao2015enquiring}\cite{ma2015detect}. Some studies investigate the aggregation of credibility from the microblog level to the event level \cite{jin2014news}. On the contrary, considering dynamic information, some work designs temporal features based on the prorogation properties over time \cite{kwon2013prominent} or trains a model with features generated from different time periods \cite{ma2015detect}. Moreover, some methods take usage of users' feedbacks (comments and attitudes) to evaluate the credibility \cite{giudice2010crowdsourcing}\cite{rieh2014audience}\cite{zhao2015enquiring}. For instance, the Enquiry Post (EP) model \cite{zhao2015enquiring} takes out signal tweets, which indicates users' suspicious attitudes for detecting rumors and achieves satisfactory performance.

It should be mentioned that the above methods have several limitations in evaluating information credibility on social media. \textit{\textbf{First}}, methods based on feature engineering usually require great labor for designing features \cite{castillo2011information}. \textit{\textbf{Secondly}}, a rough mergence resting on the statistical summation of different feature values is not competent to model complex interactions among them. For instance, there are two combinations: (1) a user with a high credibility posted a microblog and a user with low credibility reposted a microblog", (2) a user with a low credibility posted a microblog and a user with a high credibility reposted a microblog. Intuitively, the former combination is more like the style of non-rumor and the later combination is more like the style of rumor. Based on the statistical summation, the feature values of the two combinations are the same, i.e., the user credibility (high and low) and behavior type (post and repost). This rough mergence rested on simple statistical summation cannot distinguish between these two combinations. \textit{\textbf{Thirdly}}, methods based on feature engineering have difficulty in modeling some real-world scenarios from a joint perspective, e.g., who did what at when and how others reacted. Thy treat different factors (\textit{who}, \textit{what}, \textit{when}, \textit{how}) as separate features and can only extract simple compound features. For example, a user with a low credibility tended to post a rumor. There are complicated compound features, such as a user with low credibility posted a rumor at the early stage of spreading receiving suspicious comments and a user with high credibility reposted a rumor at medium term of spreading receiving identification comments. The analysis of those complicated compound features from statistical summation requires enumerating all possible compound features, which results in the explosion of time complexity and the problem of data sparsity.



\begin{figure}[tb]
\centering
\includegraphics[width=1\linewidth]{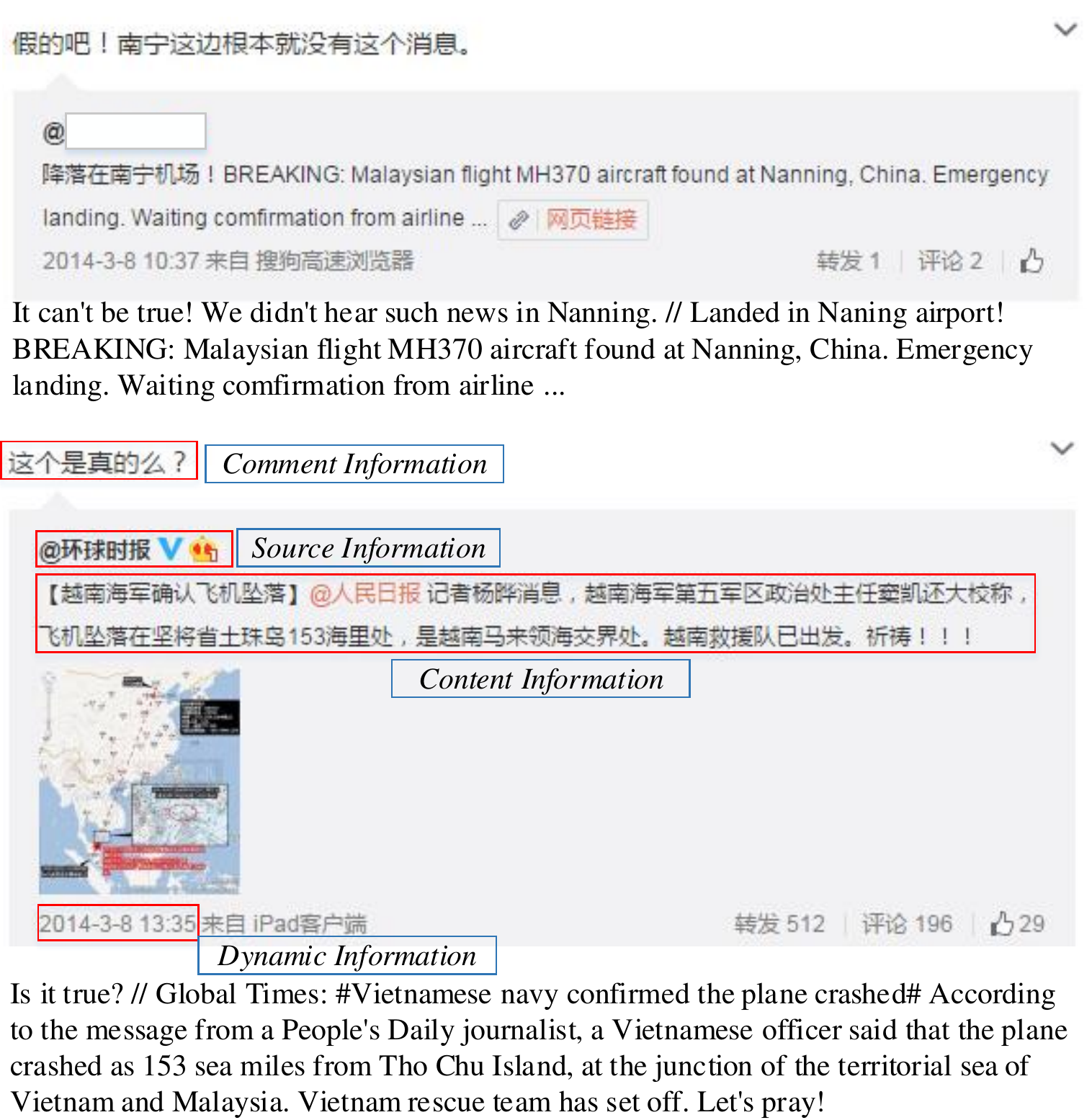}
\caption{Two rumor examples about MH370 and their repostings on Chinese Sina Weibo. The corresponding English translations are also listed here.}
\label{fig:example}
\end{figure}

In this work, we aim to evaluate the credibility of information about events on social media. Usually, each event contains several microblogs posted and reposted by users. To identify whether an event on social media is a rumor or not, we first investigate microblogs of this event. Figure \ref{fig:example} shows two examples of rumors on Sina Weibo with extracted source information, content information, temporal information, and comment information. According to such information, we conclude four key factors which are potentially useful for evaluating the information credibility on social media.

(1) \textit{\textbf{Who}} means the credibility of a user. Normally, the higher the credibility of a user, the higher the credibility of information it creates \cite{castillo2011information}. However, some studies \cite{jin2014misinformation} point out that a great amount of users with the high credibility on social media would repost and share misinformation unintentionally. As shown in the example in Figure \ref{fig:example}, even a regular media (usually with a high credibility), Global Times, would post unverified news about MH370. Therefore, it is not always reliable to model the user credibility information alone for information credibility evaluation.

(2) \textit{\textbf{What}} denotes behavior types. Usually, there are two types of behaviors for users, i.e., posting and reposting. Compared to reposts, a post indicates that the microblog is more original and relatively more important for the evaluating credibility. For rumors, original microblogs are posted by users with the relatively low credibility, whereas users with the high credibility may repost the microblogs.

\begin{figure*}[tb]
\centering
\subfigure[Microblog percentage in different time intervals since event beginning.]{
\begin{minipage}[b]{0.48\textwidth}
\centering
\includegraphics[width=0.9\textwidth]{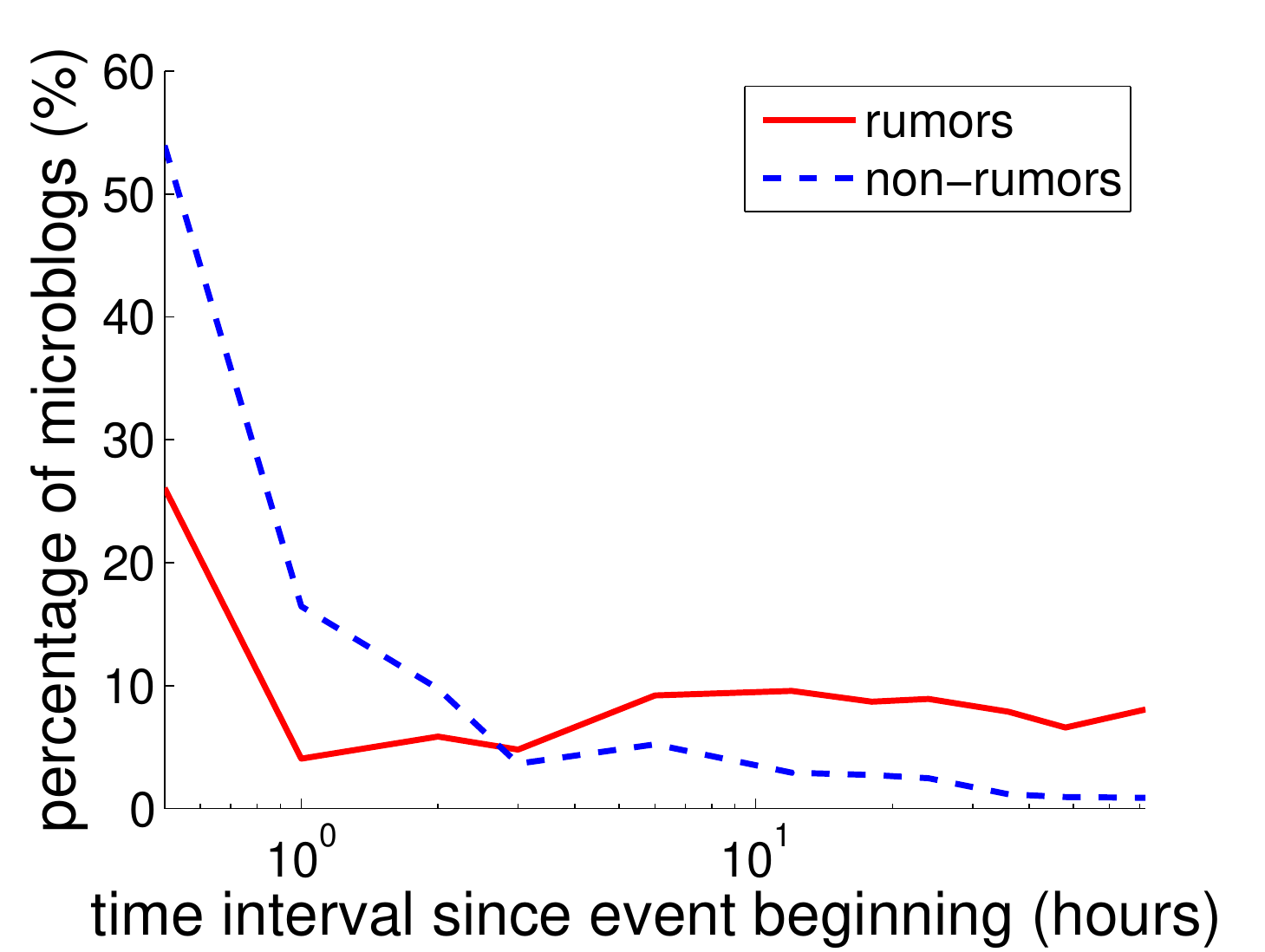}
\label{dynamic}
\end{minipage}
}
\subfigure[Distribution of percentage of suspicious microblogs of events.]{
\begin{minipage}[b]{0.48\textwidth}
\centering
\includegraphics[width=0.9\textwidth]{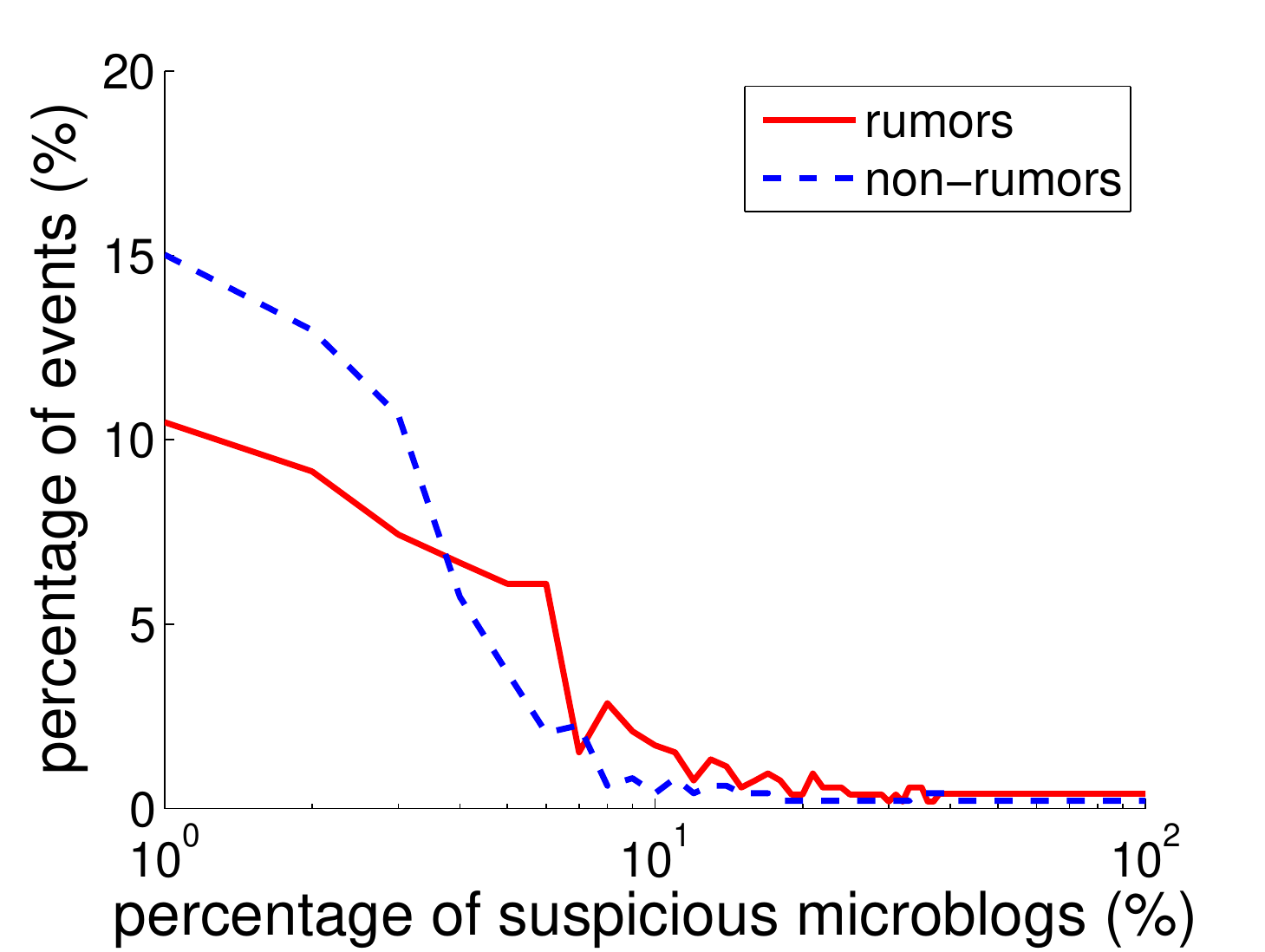}
\label{suspicious}
\end{minipage}
}
\caption{Analysis of data distribution difference between rumors and non-rumors on the Weibo dataset.}
\label{data analysis}
\end{figure*}

(3) \textit{\textbf{When}} refers to temporal properties that describe the spread process of a microblog post. As shown in Figure \ref{dynamic}, temporal properties are usually different between rumors and non-rumors. Compared to rumors, most non-rumor microblogs tend to be posted or reposted at the beginning and vanish very fast. Maybe those plain truths will become less and less attractive as time goes on. However, rumors usually draw comparatively sustained attention. The spreading curve of rumors may have multiple peaks. There might be some rumormongers promoting the spreading of rumors. In addition, for non-rumors, original microblogs at the beginning event are usually posted by users with high credibility, whereas, for rumors, original microblogs are usually posted by users with the low credibility and then reposted by other users possibly including those with the high credibility.

(4) \textit{\textbf{How}} denotes comments and attitudes towards corresponding microblogs. Users on social media can express their attitudes, and this collective intelligence can be gathered to help us evaluate the credibility of information \cite{giudice2010crowdsourcing}. Comments reveal the users' suspicion or identification attitudes towards microblogs. As shown in Figure \ref{suspicious}, rumors usually receive more suspicion comments, which is extremely helpful for detecting rumors.

The aggregation of these key factors mentioned above makes the joint perspective of a microblog and helps evaluate the credibility. However, conventional feature engineering-based methods consider these factors as separate features and roughly summarize them, which cannot well model the interactions among the key factors. We plan to learn representations jointly and obtain an overall understanding of the microblog credibility, which describes the complicated and dynamic behaviors in information spreading. This evaluation process is shaped by modeling complex interactions among different features. To be specific, we model semantic operations among different features and form an overall representation of each microblog post. Recently, representation learning \cite{bengio2013representation} is showing a promising performance in a variety of applications, such as word embedding \cite{mikolov2010recurrent}\cite{mikolov2011extensions}\cite{mikolov2013distributed}, network embedding \cite{bourigault2014learning}\cite{grover2016node}\cite{perozzi2014deepwalk}\cite{tang2015line} and user representations \cite{elkahky2015multi}\cite{feng2015personalized}\cite{liu2015collaborative}\cite{liu2015cot}.

To the end, we propose a novel ICE model that learns the joint representations of key factors of microblogs and further can be used to evaluate the credibility of events. In ICE, each user is represented as a vector according to his or her personal features, indicating the credibility information of the user (\textit{who}). For the sake of modeling complex interactions among different features, other features such as \textit{what}, \textit{how}, and \textit{when} are represented as operating matrices \cite{liu2015cot}. Behavior types (post or repost) are modeled as latent operating matrices indicating the properties of different behaviors (\textit{what}), and time intervals since the beginning of spreading of an event are represented as matrices to capture the temporal properties of behaviors (\textit{when}). Moreover, the attitudes of comments (suspicious or not) are modeled as latent operating matrices indicating collective attitudes (\textit{how}). These operating matrices model semantic operations of one feature on the others. Consequently,
the representations of dynamic and complicated behaviors can be obtained through the multiplication of user vectors with operating matrices of \textit{what}, \textit{how} and \textit{when}.
Each representation of a dynamic behavior can also be viewed as the representation of a corresponding microblog.
After aggregating all the microblog representations during information spreading, we can generate the credibility representation of the event.
Finaly, we apply a pairwise learning method to enlarge the credibility difference between rumors and non-rumors for a better and fast learning of parameters. Experiments show that our model achieves better performance compared to state-of-the-art methods.

The main contributions of this work are listed as follows:
\begin{itemize}
\item
We introduce a representation learning method for information credibility evaluation. The proposed method captures elaborate interactions among the key factors of microblogs during information spreading through learning operating matrices, which model abundant semantic operations among various features.

\item
ICE learns latent representations for the user credibility, behavior types, temporal properties and attitudes of comments. Based on these representations, ICE generates overall credibility representations of information and presents a novel perspective on information credibility evaluation.

\item
Experiments conducted on a real-world dataset show that ICE clearly outperforms the state-of-the-art methods.

\end{itemize}

The rest of the paper is organized as follows. In Section 2, we review some related work on truth discovery, credibility evaluation and representation learning. In Section 3, we introduce the used dataset and give some analysis. Section 4 details the ICE model. In Section 5, we report experimental results on the Weibo dataset and compare them to several state-of-the-art methods. In Section 6, we present a real-time information credibility evaluation system that we have constructed based on the  proposed model. Section 7 concludes our work and discusses future research.

\section{Related Work}

In this section, we briefly review some related works, including the credibility evaluation on social media, representation learning and the truth discovery.

\begin{table*}[tb]
  \centering
  \caption{Details of the Weibo dataset.}
    \begin{tabular}{cccccccc}
    \toprule
    \#events & \#rumors & \#non-rumors & \#microblogs & \#postings & \#repostings & \#users \\
    \midrule
    936   & 500   & 436   & 630,665 & 98,429 & 532,236 & 321,246 \\
    \bottomrule
    \end{tabular}%
  \label{tab:dataset}%
\end{table*}%

\subsection{Credibility Evaluation on Social Media}

Recently, some works have been proposed to automatically evaluate the information credibility and detect rumors on social media. Most of the methods are based on artificial features. Some of them evaluate the credibility of a single microblog \cite{castillo2011information}\cite{qazvinian2011rumor} or a single image \cite{gupta2013faking}. Some of them evaluate information credibility at the event level to distinguish whether an event is a rumor or a non-rumor \cite{gupta2012evaluating}\cite{sun2013detecting}\cite{kwon2013prominent}\cite{zhao2015enquiring}\cite{ma2015detect}, where each event consists of several microblogs. News Credibility Propagation (NewsCP) \cite{jin2014news} studies how to aggregate credibility from the microblog level to the event level and presents a graph optimization method, which has further incorporated conflict viewpoints in the model \cite{jin2016news}. Some works detect rumors based on the dynamic properties. For instance, the Periodic External Shocks (PES) model \cite{kwon2013prominent} uses ordinary structural features and user features and designs temporal features according to the properties of information spreading over time.The Dynamic Series-Time Structure (DSTS) \cite{ma2015detect} generates content-, user-, and diffusion-based features in different time periods during information spreading and uses all these features to train a model. Some works also take usage of users' feedbacks to evaluate credibility \cite{giudice2010crowdsourcing}\cite{rieh2014audience}. The EP model \cite{zhao2015enquiring} extracts signal tweets that indicate users' suspicious attitudes for detecting rumors and achieves satisfactory performance. The main drawback of these feature engineering-based models lies in that they require great labor for designing a great many features and cannot reveal underlying relations among these features. Moreover, these methods have difficulty in modeling elaborate interactions among different factors during information spreading.

\subsection{Representation Learning}

Nowadays, representation learning \cite{bengio2013representation} has been extensively studied in different areas. In natural language processing, learning embeddings \cite{mikolov2013distributed} is a hot topic, where recurrent neural networks \cite{mikolov2010recurrent}\cite{mikolov2011extensions} are widely applied. In web mining, learning network embedding has drawn great attention for studying node classification \cite{jacob2014learning} or information diffusion \cite{bourigault2014learning}. Recently, network embedding models have incorporated random walk \cite{perozzi2014deepwalk}\cite{grover2016node} and the second-order connection in representation learning methods \cite{tang2015line}. Meanwhile, representation models are playing a role for modeling user behaviors. Contextual Operating Tensor (COT) \cite{liu2015cot}\cite{wu2016context} and CARS2 \cite{shi2014cars2} study context-aware user representations for recommendation. Hierarchical Interaction Representation (HIR) \cite{liu2015collaborative} studies joint representations of entities, e.g., users, items and contexts, to model their interaction. Some works \cite{elkahky2015multi}\cite{wu2016auto}\cite{zhang2016deep} utilize deep neural networks for better user modeling. Convolutional Click Prediction Model (CCPM) \cite{liu2015Convolutional} applies convolutional neural networks in predicting clicking behaviors of users. Hierarchical Representation Model (HRM) \cite{wang2015learning} and Dynamic Recurrent Basket Model (DREAM)\cite{yu2016dream} learn the representation of behaviors of a user in a short period for better recommendation. These methods achieve the state-of-the-art performance in different areas, and give us inspiration for learning representations of dynamic behaviors to evaluate information credibility.

\subsection{Truth Discovery}

Truth discovery refers to the problem of finding the truth with conflicting information, which has been first addressed in \cite{yin2008truth}. It can be viewed as some kind of information credibility evaluation. Mainly based on the source credibility information, truth discovery evaluates the credibility via aggregating from different sources. Truth discovery methods are usually based on Bayesian algorithms or graph learning algorithms on stock data or flight data \cite{li2012truth}\cite{waguih2014truth}. And Semi-Supervised Truth Discovery (SSTF) \cite{yin2011semi} studies the problem with semi-supervised graph learning with a small set of ground truth data to help evaluating credibility. Truth discovery is an unsupervised or semi-supervised method to find the truth with conflicting information and make an evaluation of information credibility \cite{li2012truth}. Truth discovery is mainly based on the evaluated credibility aggregated from different information sources, usually referring to users who release information. And it is not capable to take usage of various kinds of information, such as time properties and comment attitudes, which are abundant in complex online social media scenario. Therefore, truth discovery is suitable for ideal situations with constrained topics, such as price prediction and flight arrival prediction, but hard to be applied in complex online social media.

\section{Data}

In this section, we introduce the dataset to be used in this work. Considering that there is lack of public rumor datasets, we collected a microblog dataset containing rumors and non-rumors from Sina Weibo, which is one of the biggest social media in China.

To crawl rumors, we collected some rumor seeds, i.e., some microblogs containing rumors that have been reported, from the misinformation management center of Sina Weibo. We extracted keywords from these rumor seeds and retrieved rumor microblogs with these keywords. Then, we identified the starting point of a rumor, i.e., the first microblog about the rumor, and collected all the following microblogs. For each microblog, we collected its reposting information, commenting information, and the corresponding user's profile. To crawl non-rumors, we collected some hot topics on Sina Weibo and used the same strategy as for rumors to crawl corresponding information about the non-rumors.

As shown in Table \ref{tab:dataset}, we collected $936$ events containing $500$ rumors and $436$ non-rumors. Each event consists of several microblogs (postings or repostings), and the average number is about $673$. The total number of microblogs is $630,363$, including $98,429$ postings and $532,236$ repostings. Each microblog has its posting time. The posting time of the first microblog about an event is set as the beginning of the event. The dataset contains $321,246$ users. The personal profile of a user includes gender, verified or not, number of followers, number of followees, and number of microblogs.

Moreover, considering that it is necessary to mine suspicion and identification attitudes towards microblogs from comments, we need to annotate each microblog being  suspicious or not. For there is no proper corpus for training a classifier about suspicion, we used an unsupervised method to identify suspicious attitudes towards microblogs. We built up a list of suspicion words and distinguished a microblog according to whether those suspicion words appear in the microblog. We first found several typical suspicion words then train wrod2vec\footnote{https://code.google.com/p/word2vec/} \cite{mikolov2013distributed} on our dataset and found dozens of words similar with the typical suspicion words according to their embedding distance. Finally, we built up a word list with about 100 suspicion words and annotated all the microblogs being suspicious or not in our Weibo dataset.

Based on the dataset, we also investigated the data distribution difference between rumors and non-rumors, which is shown in Figure \ref{data analysis}, i.e., distribution of percentage of microblogs with time illustrated in Figure \ref{dynamic} and distribution of percentage of events with the percentage of suspicious microblogs in one event shown in Figure \ref{suspicious}.

\section{The ICE model}

In this section, we first formulate the problem. Then, we detail the proposed ICE model. Finally, we present the pairwise learning procedure for the ICE model.

\subsection{Problem Formulation}
The problem to be studied in this paper can be formulated as follows. Suppose a set of events are denoted as $E = \left\{ {e_1 ,e_2 ,...,e_n } \right\}$ and $s_{e_i}$ is the credibility score of the corresponding event $e_i$. $s_{e_i} = 0$ means event $e_i $ is a rumor and $s_{e_i }  = 1$ means event $e_i $ is a non-rumor. The microblogs of the event $e_i$ can be denoted as $M^{e_i }  = \left\{ {m_1^{e_i } ,m_2^{e_i } ,...,m_{n_{e_i}}^{e_i }} \right\}$, where $n_{e_i}$ is the number of microblogs of this event. All microblog sets can be written as $M = \left\{ {M^{e_1 } ,M^{e_2 } ,...,M^{e_n } } \right\}$. Each microblog $m_j^{e_i }$ consists of four elements \textit{\textbf{who}}, \textit{\textbf{what}}, \textit{\textbf{when}} and \textit{\textbf{how}}, which are denoted as $u_j^{e_i }$, $b_j^{e_i }$, $c_j^{e_i }$ and $t_j^{e_i }$. $u_j^{e_i }$ is the corresponding user of the microblog, $b_j^{e_i }$ denotes the behavior type (posting or reposting), $c_j^{e_i }$ describes the user's comments and attitudes (suspicious or not), and $t_j^{e_i }$ denotes the time interval since the beginning of an event. In this work, our task is to evaluate the credibility of an event on social media.

\subsection{Proposed Model}

Here, we detail the representation learning procedure of the ICE model. Based on handcrafted features that indicate global-wise statistics, conventional methods have difficulty in effectively modeling the correlation among different key elements in information spreading. Thus, we need to model their joint representations and yield their joint characteristics. It is necessary for a model based on the user credibility (who), behavior types (what), comment attitudes (how) and dynamic properties (when).

We first start with user information and behavior information. User information tells us the properties and the credibility of a user. Behavior information tells us the behavior type, i.e., posting or reposing. Moreover, the combination of user and behavior information shows who did what. Mathematically, for the $j$-th microblog $m_j^{e_i }$ of the event $e_i$, the representation of this microblog with the user $u_j^{e_i }$ and the behavior $b_j^{e_i }$ can be written as

\begin{equation} \label{MR1}
{\mathbf{R}}_j^{e_i }  = {\mathbf{B}}_j^{e_i} {\mathbf{U}}_j^{e_i} ~,
\end{equation}
where ${\mathbf{U}}_j^{e_i} \in \mathbb{R}^{d}$ is the vector representation of user $u_j^{e_i }$, ${\mathbf{B}}_j^{e_i} \in \mathbb{R}^{d \times d}$ is the matrix representation of behavior $b_j^{e_i }$, and $d$ denotes the dimensionality of representations.

Additionally, users may express their attitudes in the comments. These attitudes contain the knowledge and life experience of users and can be used to distinguish rumors from non-rumors. As shown in Figure \ref{suspicious}, rumors often receive more suspicious comments than non-rumors. Incorporating the representation of comment attitude $c_j^{e_i }$ of microblog $m_j^{e_i }$, Equation \ref{MR1} can be further written as
\begin{equation} \label{MR2}
{\mathbf{R}}_j^{e_i }  = {\mathbf{C}}_j^{e_i} {\mathbf{B}}_j^{e_i} {\mathbf{U}}_j^{e_i} ~,
\end{equation}
where ${\mathbf{C}}_j^{e_i} \in \mathbb{R}^{d \times d}$ is the matrix representation of the comment $c_j^{e_i }$. Now, this equation can reveal the joint representation of who did what under how.

\begin{figure}[tb]
\centering
\includegraphics[width=1\linewidth]{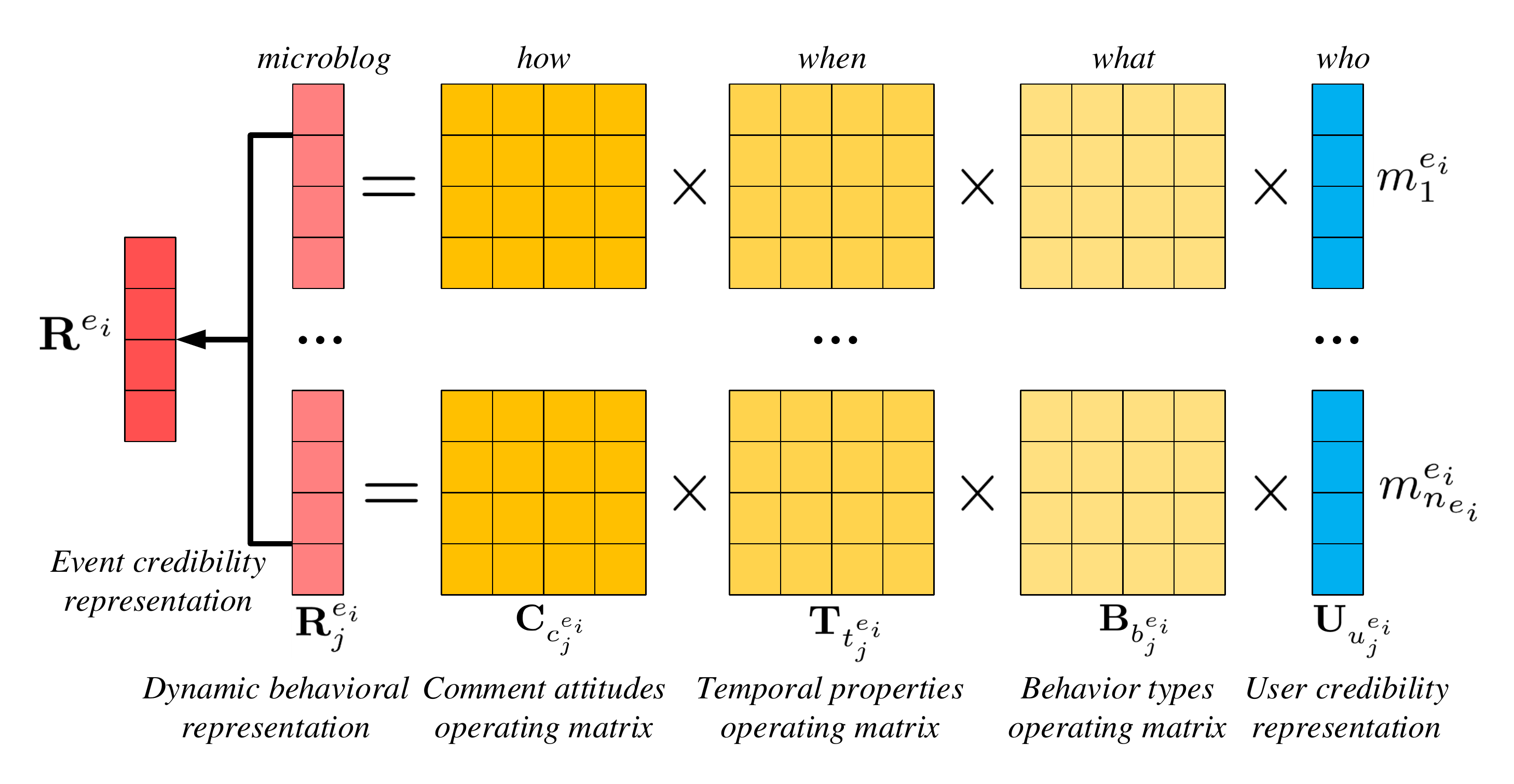}
\caption{Overview of the representation learning procedure in the proposed ICE model.}
\label{fig:model}
\end{figure}

Moreover, Figure \ref{dynamic} illustrates the difference between dynamic properties of rumors and non-rumors. It shows that time interval information is a significant factor for evaluating the information credibility and should be modeled jointly with user behaviors.
For instance, time interval $t_j^{e_i }$ of microblog $m_j^{e_i}$ means that the microblog appears from the beginning of $e_i$.
Incorporating time interval $t_j^{e_i }$ in Equation \ref{MR2}, the representation of microblog $m_j^{e_i }$ can be rewritten as
\begin{equation} \label{MR}
{\mathbf{R}}_j^{e_i }  = {\mathbf{T}}_j^{e_i} {\mathbf{C}}_j^{e_i} {\mathbf{B}}_j^{e_i} {\mathbf{U}}_j^{e_i} ~,
\end{equation}
where ${\mathbf{T}}_j^{e_i} \in \mathbb{R}^{d \times d}$ is the matrix representation of time interval $t_j^{e_i }$.
Now, this equation can reveal the joint representation of who did what under how at when.

Right now, we generate the representation ${\mathbf{R}}_j^{e_i}$ of the microblog $m_j^{e_i}$, which can capture the joint properties of four key elements. Because each event consists of several microblogs, we need to aggregate all representations of microblogs to generate the final credibility representation of the event. Using the average calculation, the representation of the event $e_i$ can be generated as
\begin{equation} \label{ER}
\begin{split}
{\mathbf{R}}^{e_i }  = \frac{1}{{n_{e_i } }}\sum\limits_{m_j^{e_i }  \in M^{e_i } } {{\mathbf{R}}_j^{e_i } }~~~~~~~~~~~~~~~~~~~~ \\
= \frac{1}{{n_{e_i } }}\sum\limits_{m_j^{e_i }  \in M^{e_i } } {{\mathbf{T}}_j^{e_i} {\mathbf{C}}_j^{e_i} {\mathbf{B}}_j^{e_i} {\mathbf{U}}_j^{e_i } } ~.
\end{split}
\end{equation}

Then, we can predict whether an event $e_i$ is a rumor or not using
\begin{equation} \label{RP}
y^{e_i }  = {\mathbf{W}}^T {\mathbf{R}}^{e_i } ~,
\end{equation}
where ${\mathbf{W}} \in \mathbb{R}^{d}$ is linear weights of the prediction function. A larger value of $y^{e_i }$ indicates a higher credibility of $e_i$.

\subsection{User Representation Generation}

\begin{figure}[tb]
\centering
\includegraphics[width=1.05\linewidth]{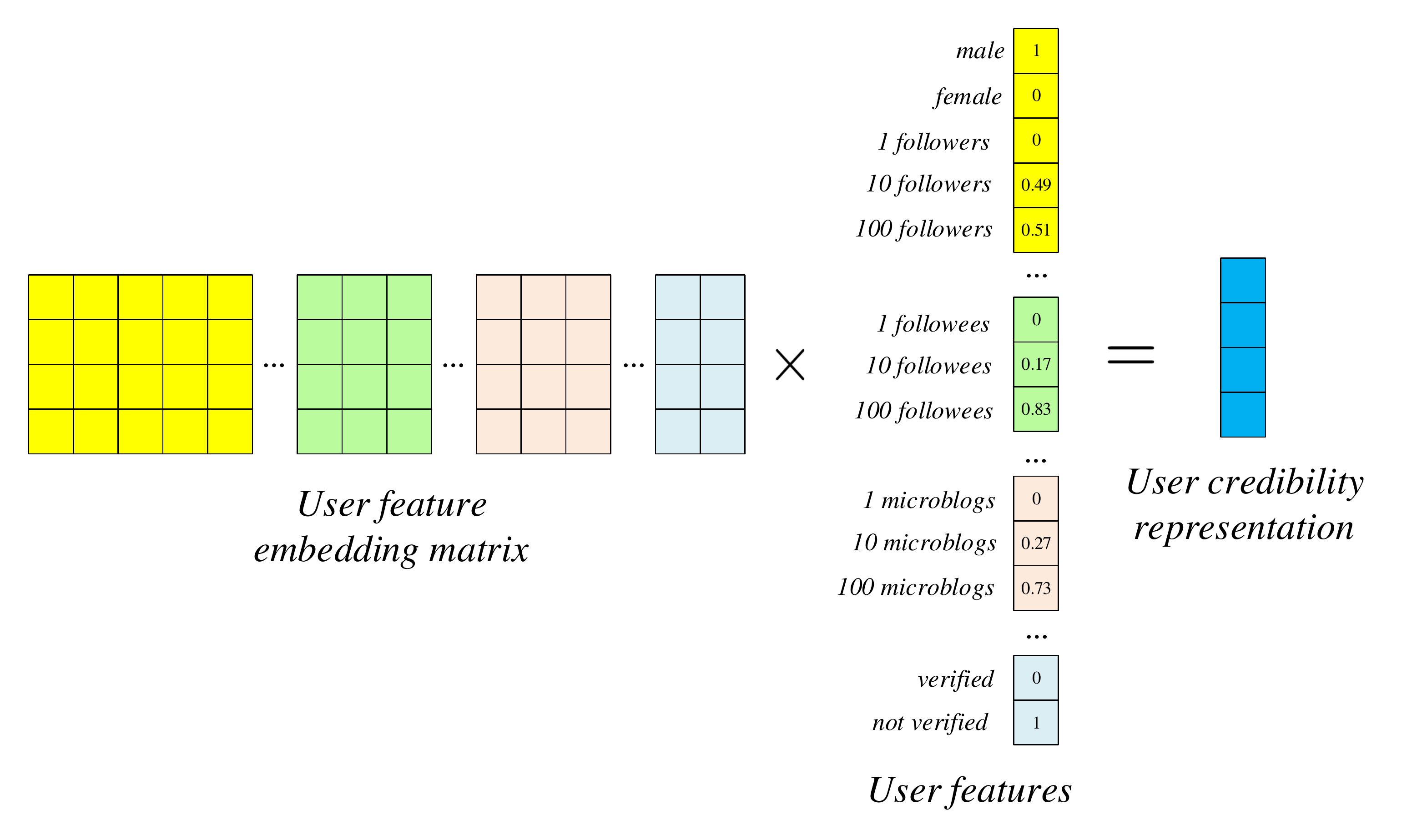}
\caption{An example of generating the user representation. The user's features are \{male, $32$ followers, $68$ followees, $54$ microblogs, not verified\}.}
\label{fig:user}
\end{figure}

For learning user representations, i.e., who, in the ICE model, it would be desirable if we can learn a distinct latent vector for each user to capture his or her properties and credibility. However, according to Table \ref{tab:dataset}, each user (re)posts only two microblogs in average, which cannot bring enough information to directly learn a latent representation for each user.

Instead, we can learn the embeddings of rich features for users. These features contained in the Weibo dataset are gender, number of followers, number of followees, numbers of microblogs, and verified or not. Accordingly, users can be shaped based on the above features. For user $u$, we have a feature vector ${\mathbf{F}_u} \in \mathbb{R}^{f}$ which is constructed as
\begin{displaymath}
{\mathbf{F}_u} =  [{\mathbf{F}^{gender}_u},{\mathbf{F}^{followers}_u},{\mathbf{F}^{followees}_u},{\mathbf{F}^{microblogs}_u},{\mathbf{F}^{verified}_u}]^T~.
\end{displaymath}
Both ${\mathbf{F}^{gender}_u}$ and ${\mathbf{F}^{verified}_u}$ have two bits. ${\mathbf{F}^{gender}_u}(1)=1$ denotes that the gender is male, and ${\mathbf{F}^{gender}_u}(2)=1$ denotes that the gender is female. ${\mathbf{F}^{verified}_u}(1)=1$ means that the user is verified, and ${\mathbf{F}^{verified}_u}(2)=1$ otherwise. For the numbers of followers, followees, and microblogs, it is hard to learn an embedding for each distinct value. Therefore, we partition the values into the discrete bins according to a $log_{10}$ distribution. If a user $u$ has $v_u$ followers, the corresponding features can be constructed as
\begin{displaymath}
{\mathbf{F}^{followers}_u}(i)=\left\{
\begin{aligned}
U(log^{v_u}_{10})-log^{v_u}_{10},i={L(log^{v_u}_{10})}+1 \\
log^{v_u}_{10}-L(log^{v_u}_{10}),i={U(log^{v_u}_{10})}+1 \\
0~~~~~~~~~~,i=others~~~~~~~~
\end{aligned}
\right.~,
\end{displaymath}
where $U(log^{v_u}_{10})$ and $L(log^{v_u}_{10})$ denote the upper and lower bounds of $log^{v_u}_{10}$ respectively. Meanwhile, ${\mathbf{F}^{followees}_u}$ and ${\mathbf{F}^{microblogs}_u}$ can be constructed in the same way. Figure \ref{fig:user} illustrates an example of generating the user representation, in which we suppose $v_u = 32$, then $log^{32}_{10} = 1.51$, the corresponding upper and lower bounds are $2$ and $1$. ${\mathbf{F}^{followers}_u}$ can be computed as
\begin{displaymath}
{\mathbf{F}^{followers}_u}(2)=2-1.51=0.49 ~,
\end{displaymath}
\begin{displaymath}
{\mathbf{F}^{followers}_u}(3)=1.51-1=0.51 ~,
\end{displaymath}
and other bits will be set to be $0$.

Then, based on the feature vector ${\mathbf{F}_u}$, we can generate the user representation as
\begin{equation}
{\mathbf{U}}_u  = {\mathbf{S}} {\mathbf{F}_u} ~,
\end{equation}
where ${\mathbf{S}} \in \mathbb{R}^{d \times f}$ is the feature embedding matrix.

\subsection{Nonlinear Interpolation for Generating Time-Specific Matrices}

In ICE, we use time-specific matrices to capture the properties of users' dynamic behaviors, i.e., when, in the information spreading. However, if we learn a distinct matrix for each possible continuous time interval, the ICE model will face the data sparsity problem. Therefore, as in \cite{liu2016strnn}, we use a similar strategy for generating time-specific matrices. We partition the time interval into discrete time bins. Considering the power law distribution of dynamic behaviors shown in Figure \ref{dynamic}, it is not plausible if we partition the time interval equally. Instead, our partition confirms to a $log_2$ distribution. Only the matrices of the upper and lower bounds of the corresponding bins are learned in our model. For time intervals in a time bin, their transition matrices can be calculated via a nonlinear interpolation.

\begin{figure}[tb]
\centering
\includegraphics[width=1\linewidth]{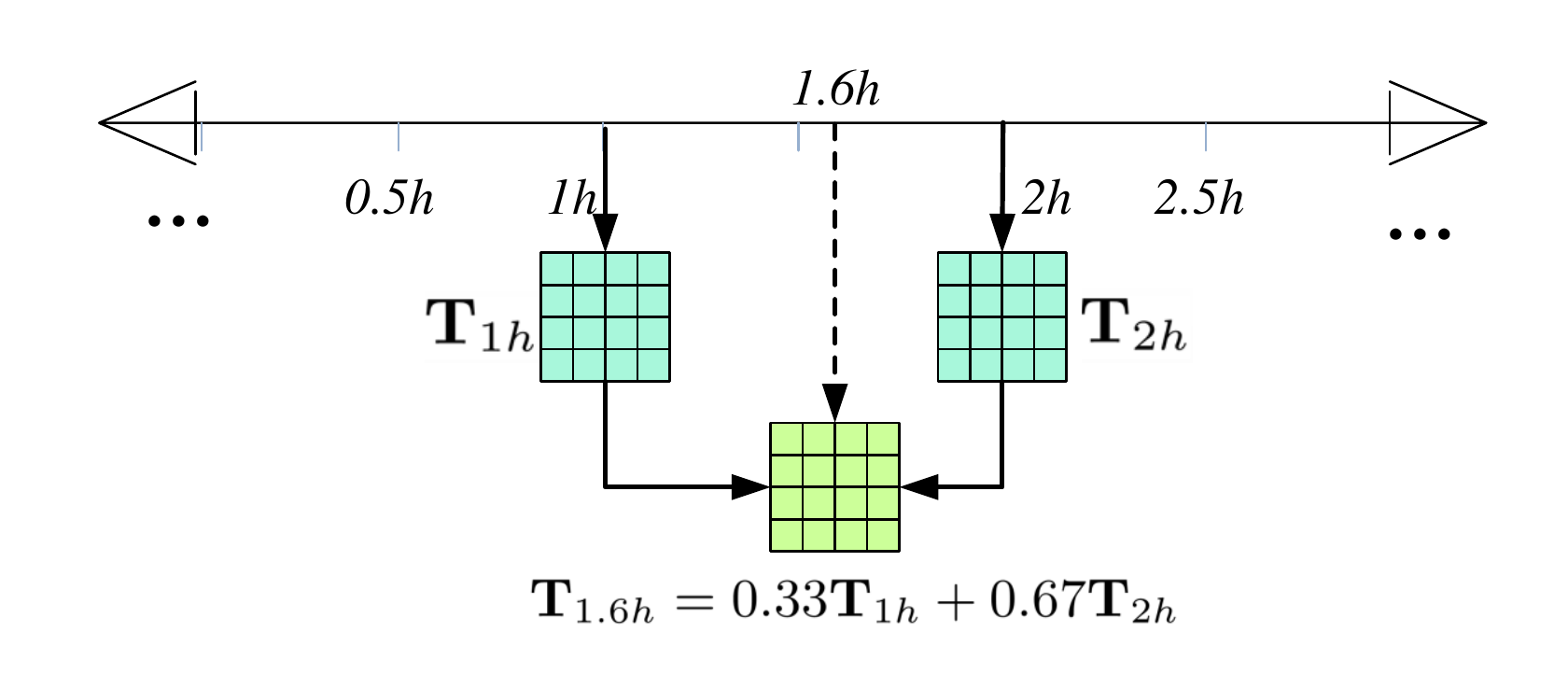}
\caption{An example of generating the time-specific matrix. The time interval in this example is $1.6h$. Via a nonlinear interpolation of $1h$ and $2h$, the corresponding time-specific matrix can be generated as $\mathbf{T}_{1.6h} = 0.33\mathbf{T}_{1h}+0.67\mathbf{T}_{2h}$.}
\label{fig:time}
\end{figure}

Mathematically, the time-specific matrix $\mathbf{T}_t$ for time interval $t$ can be calculated as
\begin{displaymath}
\mathbf{T}_{t}  = \frac{{ {(U(log^t_2) - log^t_2) \mathbf{T}_{2^{L(log^t_2)}} + (log^t_2 - L(log^t_2)) \mathbf{T}_{2^{U(log^t_2)}}} }}{{{U(log^t_2) -  L(log^t_2)}}} ~,
\end{displaymath}
where $U(log^t_2)$ and $L(log^t_2)$ denote the upper bound and lower bounds of $log^t_2$ respectively. An example is shown is Figure \ref{fig:time}, where $t=1.6h$, $log^{1.6h}_2 = 0.67$, the corresponding upper and lower bounds will be $1$ and $0$, respectively; then, $\mathbf{T}_{1.6h}$ can be computed as
\begin{displaymath}
\begin{aligned}
&\mathbf{T}_{1.6h} = \frac{{ {(1 - 0.67) \mathbf{T}_{1h} + (0.67 - 0) \mathbf{T}_{2h}} }}{{{1 -  0}}} \\
&~~~~~~~= 0.33\mathbf{T}_{1h}+0.67\mathbf{T}_{2h} \\
 \end{aligned}~.
\end{displaymath}
Such an interpolation method can solve the problem of learning matrices for continuous values in the ICE model and provides a solution for modeling the dynamic behaviors of users.

\subsection{Pair-wise Learning}

Here, we introduce the parameter estimation process of ICE with a pair-wise learning method and calculate the complexity of the algorithm.

Because rumors are often hard to collect for training credibility evaluation models, we apply a pair-wise learning method to enlarge the number of training instances. Similar to \cite{rendle2009bpr}, our basic assumption is that the credibility of a non-rumor is larger than that of a rumor. In ICE, we can maximize the credibility difference between rumors and non-rumors. Accordingly, we should maximize the following probability:
\begin{displaymath}
p(e_n \succ e_{r}) = g(y^{e_n }  - y^{e_r } ) ~,
\end{displaymath}
where $e_n$ denotes a non-rumor, $e_r$ denotes a rumor, and $g(x)$ is a nonlinear function that is selected as:
\begin{displaymath}
g(x) = \frac{1}{{1 + e^{ - x} }} ~.
\end{displaymath}
Incorporating the negative log likelihood, for the whole dataset, we can minimize the following objective function equivalently:
\begin{displaymath}
J = \sum\limits_{\{e_n,e_r \} \in E,l_{e_n}=1,l_{e_r}=0} {\ln (1 + e^{ - {\mathbf{W}}^T ({\mathbf{R}}^{e_n }  - {\mathbf{R}}^{e_r }) } )}  + \frac{\lambda }{2}\left\| \mathbf{\Theta}  \right\|^2 ~,
\end{displaymath}
where $\mathbf{\Theta}  = \left\{ {\mathbf{U},\mathbf{B},\mathbf{C},\mathbf{T},\mathbf{W}} \right\}$ denotes all the parameters to be estimated and $\lambda $ is a parameter to control the power of regularization. The derivations of $J$ with respect to $\mathbf{W}$, ${\mathbf{R}}^{e_n }$ and ${\mathbf{R}}^{e_r }$ can be calculated as
\begin{displaymath}
\frac{{\partial J}}{{\partial \mathbf{W} }} = \sum\limits_{e_n,e_r \in E,l_{e_n}=1,l_{e_r}=0} {\frac{{({\mathbf{R}}^{e_r }  - {\mathbf{R}}^{e_n }) l(e_n,e_r) }}{{1 + l(e_n,e_r) }}}  + \lambda \mathbf{W}  ~,
\end{displaymath}
\begin{displaymath}
\frac{{\partial J}}{{\partial {\mathbf{R}}^{e_n } }} = - \sum\limits_{e_r \in E,l_{e_r}=0} {\frac{{\mathbf{W} l(e_n,e_r) }}{{1 + l(e_n,e_r) }}}  ~,
\end{displaymath}
\begin{displaymath}
\frac{{\partial J}}{{\partial {\mathbf{R}}^{e_r } }} = \sum\limits_{e_n \in E,l_{e_n}=1} {\frac{{\mathbf{W} l(e_n,e_r) }}{{1 + l(e_n,e_r) }}}  ~.
\end{displaymath}
where
\begin{displaymath}
l(e_n,e_r) = e^{ - {\mathbf{W}}^T ({\mathbf{R}}^{e_n }  - {\mathbf{R}}^{e_r }) }  ~.
\end{displaymath}

After calculating the derivation ${\partial J \mathord{\left/ \right. \kern-\nulldelimiterspace} {\partial {\mathbf{R}}^{e_i } } }$ of the event representation ${\mathbf{R}}^{e_i }$, the corresponding gradients of all the parameters can be calculated as
\begin{displaymath}
\frac{{\partial J}}{{\partial {\mathbf{T}}_j^{e_i }}} = \frac{1}{{n_{e_i } }} \frac{{\partial J}}{{\partial {\mathbf{R}}^{e_i } }} \left({\mathbf{C}}_j^{e_i} {\mathbf{B}}_j^{e_i} {\mathbf{U}}_j^{e_i}  \right)^T ~,
\end{displaymath}
\begin{displaymath}
\frac{{\partial J}}{{\partial {\mathbf{C}}_j^{e_i} }} = \frac{1}{{n_{e_i } }} \left( {\mathbf{T}}_j^{e_i} \right)^T \frac{{\partial J}}{{\partial {\mathbf{R}}^{e_i } }} \left({\mathbf{B}}_j^{e_i} {\mathbf{U}}_j^{e_i}  \right)^T ~,
\end{displaymath}
\begin{displaymath}
\frac{{\partial J}}{{\partial {\mathbf{B}}_j^{e_i }}} = \frac{1}{{n_{e_i } }} \left( {\mathbf{T}}_j^{e_i} {\mathbf{C}}_j^{e_i} \right)^T \frac{{\partial J}}{{\partial {\mathbf{R}}^{e_i } }} \left({\mathbf{U}}_j^{e_i}  \right)^T ~,
\end{displaymath}
\begin{displaymath}
\frac{{\partial J}}{{\partial {\mathbf{U}}_j^{e_i }}} = \frac{1}{{n_{e_i } }} \left( {\mathbf{T}}_j^{e_i} {\mathbf{C}}_j^{e_i} {\mathbf{B}}_j^{e_i } \right)^T \frac{{\partial J}}{{\partial {\mathbf{R}}^{e_i } }} ~.
\end{displaymath}
Then,
\begin{displaymath}
\frac{{\partial J}}{{\partial {\mathbf{T}}_{t} }} = \sum\limits_{e_i \in E} \sum\limits_{m_j^{e_i } \in M^{e_i },t_j^{e_i }=t} \frac{{\partial J}}{{\partial {\mathbf{T}}_j^{e_i }}} +\lambda {\mathbf{T}}_{t}  ~,
\end{displaymath}
\begin{displaymath}
\frac{{\partial J}}{{\partial {\mathbf{C}}_{c} }} = \sum\limits_{e_i \in E} \sum\limits_{m_j^{e_i } \in M^{e_i },c_j^{e_i }=c} \frac{{\partial J}}{{\partial {\mathbf{C}}_j^{e_i }}} +\lambda {\mathbf{C}}_{c}  ~,
\end{displaymath}
\begin{displaymath}
\frac{{\partial J}}{{\partial {\mathbf{B}}_{b} }} = \sum\limits_{e_i \in E} \sum\limits_{m_j^{e_i } \in M^{e_i },b_j^{e_i }=b} \frac{{\partial J}}{{\partial {\mathbf{B}}_j^{e_i }}} +\lambda {\mathbf{B}}_{b}  ~,
\end{displaymath}
\begin{displaymath}
\frac{{\partial J}}{{\partial {\mathbf{U}}_{u} }} = \sum\limits_{e_i \in E} \sum\limits_{m_j^{e_i } \in M^{e_i },u_j^{e_i }=u} \frac{{\partial J}}{{\partial {\mathbf{U}}_j^{e_i }}} +\lambda {\mathbf{U}}_{u}  ~.
\end{displaymath}
After all the gradients are calculated, we can employ the gradient descent to estimate the model parameters. This process can be repeated iteratively until convergence.

Based on the above calculation, now we analyze the corresponding time complexity and suppose we totally have $n$ events with $m$ microblogs. During the training procedure, in each iteration, the time complexities of updating $T$, $C$, $B$ and $U$ are $O(d^2 \times m)$ respectively. And the time complexity of updating $W$ is $O(d \times n)$. So, the total time complexity is $O[4d^2 \times m + d \times n]$. Since $m$ is usually much larger than $n$ and $d$ is a constant, the time complexity is approximately equal to $O(m)$. During the testing procedure, the time complexity is $O(d^2 \times m + d \times n)$. It is also approximately equal to $O(m)$. This indicates that both training and testing time complexities grow linearly with the size of the dataset, and ICE has a potential to be scaled up to large-scale data.

\section{Experiments}

In this section, we conduct empirical experiments to demonstrate the effectiveness of the ICE model on the Sina Weibo dataset. We first introduce settings of our experiments. Then, we compare the ICE model to the state-of-the-art baseline methods. We also study the performance of the ICE model with varying parameters and under different situations. Finally, we analyze the scalability of the ICE model.

\begin{table*}[tb]
  \centering
  \caption{Performance comparison evaluated with the dimensionality $d=8$.}
    \begin{tabular}{c|c|ccc|ccc}
    \toprule
    \multirow{2}[0]{*}{Methods} & \multirow{2}[0]{*}{Accuracy} & \multicolumn{3}{c|}{Rumors} & \multicolumn{3}{c}{Non-rumors} \\
          &       & Precision & Recall & F1-score & Precision & Recall & F1-score \\
    \midrule
    NewsCP-Content & 0.608  & 0.524  & 0.899  & 0.662  & 0.531  & 0.782  & 0.633  \\
    NewsCP-Social & 0.618  & 0.617  & 0.929  & 0.742  & 0.556  & 0.793  & 0.654  \\
    NewsCP & 0.758  & 0.741  & 0.808  & 0.773  & 0.728  & 0.770  & 0.749  \\
    EP    & 0.812  & 0.795  & 0.899  & 0.844  & 0.802  & 0.793  & 0.798  \\
    EP+Content & 0.823  & 0.809  & 0.899  & 0.852  & 0.868  & 0.759  & 0.810  \\
    \textbf{ICE}   & 0.860  & 0.830  & 0.939  & 0.882  & 0.919  & 0.782  & 0.845  \\
    \textbf{ICE+Content} & \textbf{0.887} & \textbf{0.831} & \textbf{0.990} & \textbf{0.903} & \textbf{0.946} & \textbf{0.805} & \textbf{0.870} \\
    \bottomrule
    \end{tabular}%
  \label{tab:performance}%
\end{table*}%

\subsection{Experimental Settings}

First, we split our dataset into the training set and testing set. Randomly, we use $70\%$ of the events (rumors or non-rumors) in the dataset for training, and the remaining $30\%$ for testing.

Moreover, we choose several evaluation metrics for our experiments: \textbf{Accuracy}, \textbf{Precision}, \textbf{Recall}, and \textbf{F1-score}. $Accuracy$ is a standard metric for classification tasks, which is evaluated by the percentage of correctly predicted rumors and non-rumors. $Precision$, $Recall$ and $F1$-$score$ are widely-used metrics for classification tasks, which are computed according to where correctly predicted rumors or non-rumors appear in the predicted list. The larger the values of the above evaluation metrics, the better the performance.

Three competitive methods and their extensions are compared in our experiments:
\begin{itemize}
\item News Credibility Propagation (NewsCP) \cite{jin2014news} studies how to aggregate credibility from microblogs to events based on a graph optimization method. The classifier we use for each microblog is the widely-used Support Vector Machine (SVM), which is implemented via libSVM\footnote{http://www.csie.ntu.edu.tw/~cjlin/libsvm/index.html} \cite{chang2011libsvm}. There are three different versions of NewsCP: \textbf{NewsCP-Content}, \textbf{NewsCP-Social}, and \textbf{NewsCP}. NewsCP-Content only uses the content of microblogs as features. Meanwhile, NewsCP-Social only uses social features of the corresponding microblogs. NewsCP takes usage of all the features. Social features include number of user followers, number of user followees, number of user microblogs, gender of user, user verified or not, number of repostings, number of comments and time of posting.

\item The Enquiry Post (EP) model \cite{zhao2015enquiring} is proposed mainly based on signal tweets. We use our suspicion word list to identify signal tweets and then apply libSVM \cite{chang2011libsvm} for information credibility evaluation. The features used here include the percentage of signal tweets, content length, average number of repostings, average number of URLs, average number of hashtags, average number of usernames mentioned, and average time of posting. Considering that \textbf{EP} does not take content information of microblogs into consideration, we further make a fusion of EP and NewsCP-Content at the score level and achieve an extended version \textbf{EP+Content}.

\item Our proposed \textbf{ICE} model uses the representation learning method to the evaluate credibility. Similar to the EP model, considering ICE only models user behaviors, we make a fusion of ICE and NewsCP-Content at the score level to incorporate content information and achieve an extended version \textbf{ICE+Content}.
\end{itemize}
Note that, the score level fusion means the final predicted score is the sum of the predicted scores of the two methods. Mathematically, for fusing the scores of ICE and NewsCP-Content to generate the score of ICE+Content, it can be calculated as:
\begin{displaymath}
S_{ICE+Content} = \mu S_{ICE} + (1-\mu) S_{NewsCP-Content} ~,
\end{displaymath}
where $S_{ICE+Content}$, $S_{ICE}$, and $S_{NewsCP-Content}$ denote predicted credibility scores of methods ICE+Content, ICE, and NewsCP-Content respectively, and $\mu$ is empirically selected to be $\mu  = 0.8$ in our experiments. For generating the predicted credibility score of EP+Content, the process is the same.

\begin{figure*}[tb]
\centering
\subfigure[$Precision-Recall$ curves for rumors.]{
\begin{minipage}[b]{0.48\textwidth}
\centering
\includegraphics[width=0.9\textwidth]{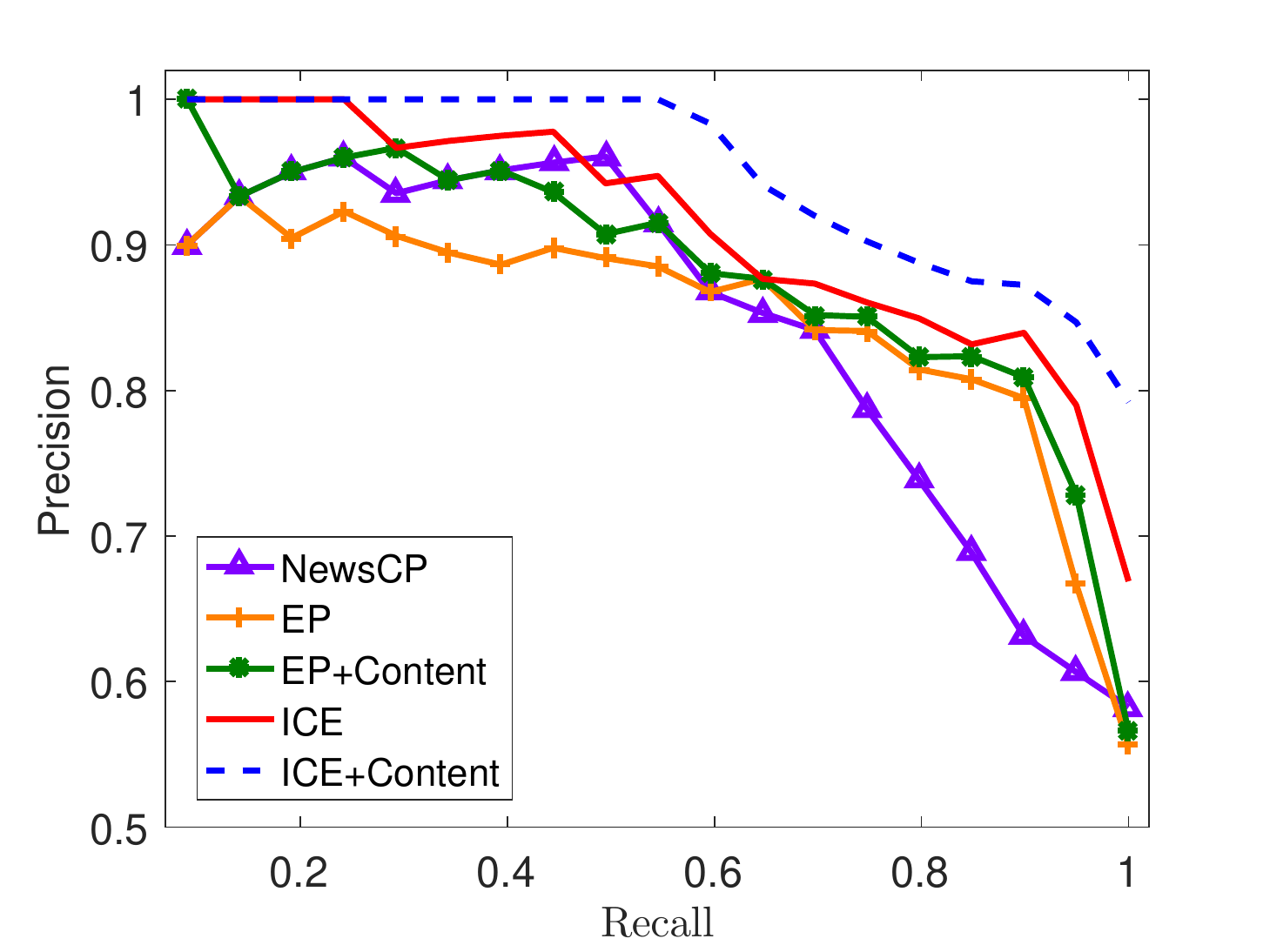}
\label{fig:PR1}
\end{minipage}
}
\subfigure[$Precision-Recall$ curves for non-rumors.]{
\begin{minipage}[b]{0.48\textwidth}
\centering
\includegraphics[width=0.9\textwidth]{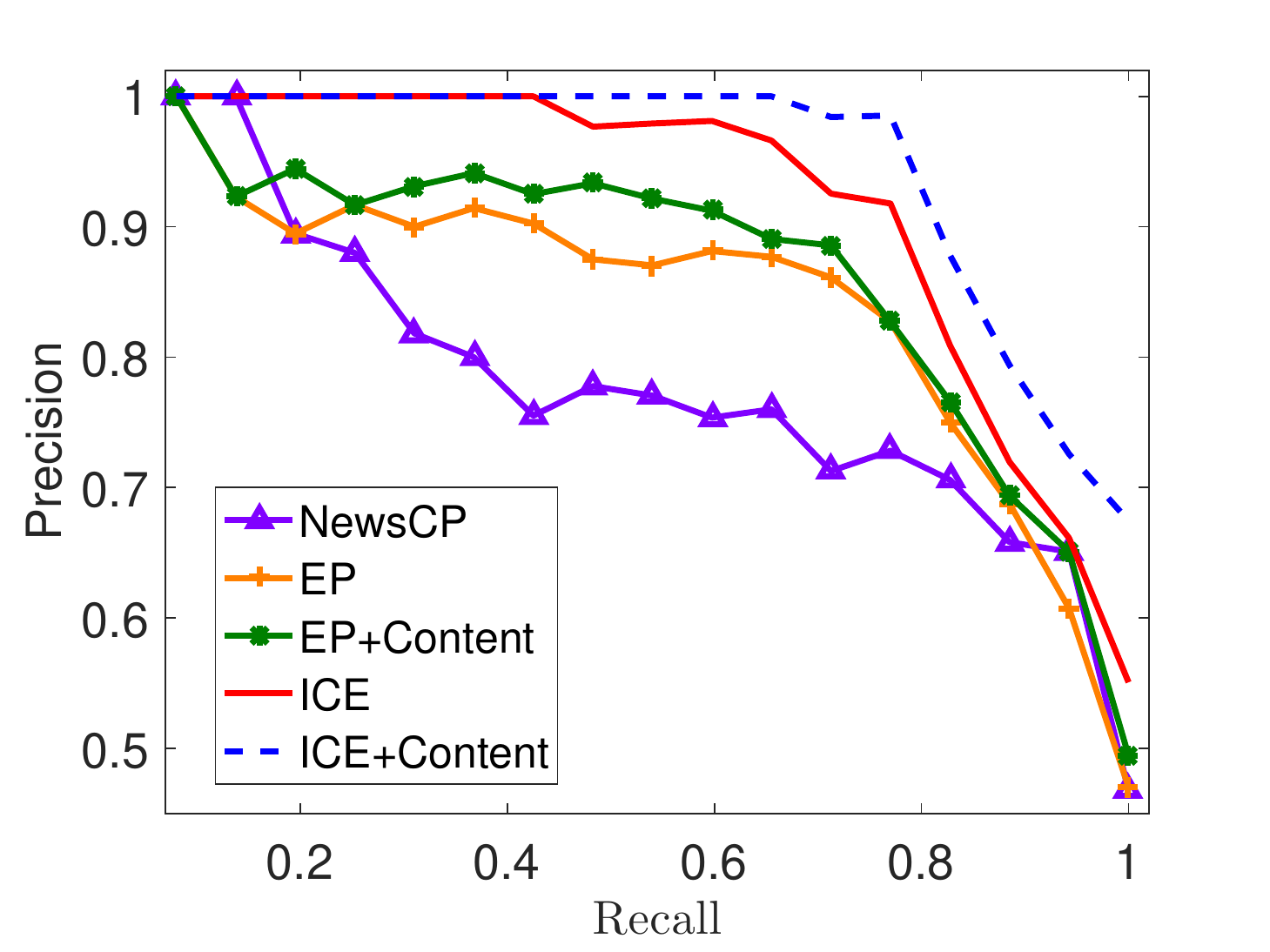}
\label{fig:PR2}
\end{minipage}
}
\caption{$Precision$-$Recall$ curves of different methods with dimensionality $d=8$.}
\label{fig:PR}
\end{figure*}

\begin{figure*}[tb]
\centering
\subfigure[$Accuracy$ curves of ICE with varying dimensionality $d$ with $\lambda=0.01$.]{
\begin{minipage}[b]{0.48\textwidth}
\centering
\includegraphics[width=0.9\textwidth]{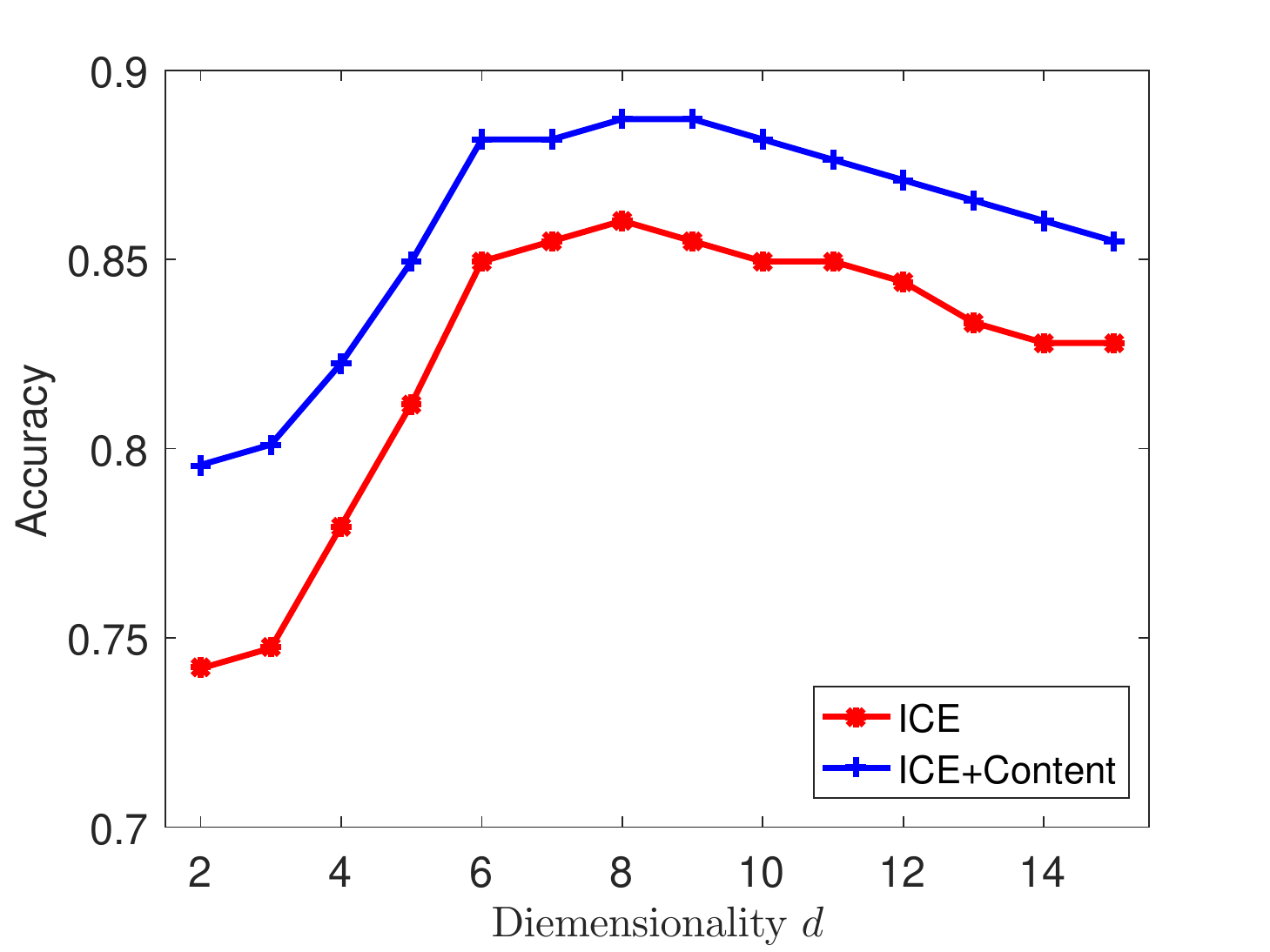}
\label{fig:dimensionality}
\end{minipage}
}
\subfigure[$Accuracy$ curves of ICE with varying regularization parameter $\lambda$ with $d=8$.]{
\begin{minipage}[b]{0.48\textwidth}
\centering
\includegraphics[width=0.9\textwidth]{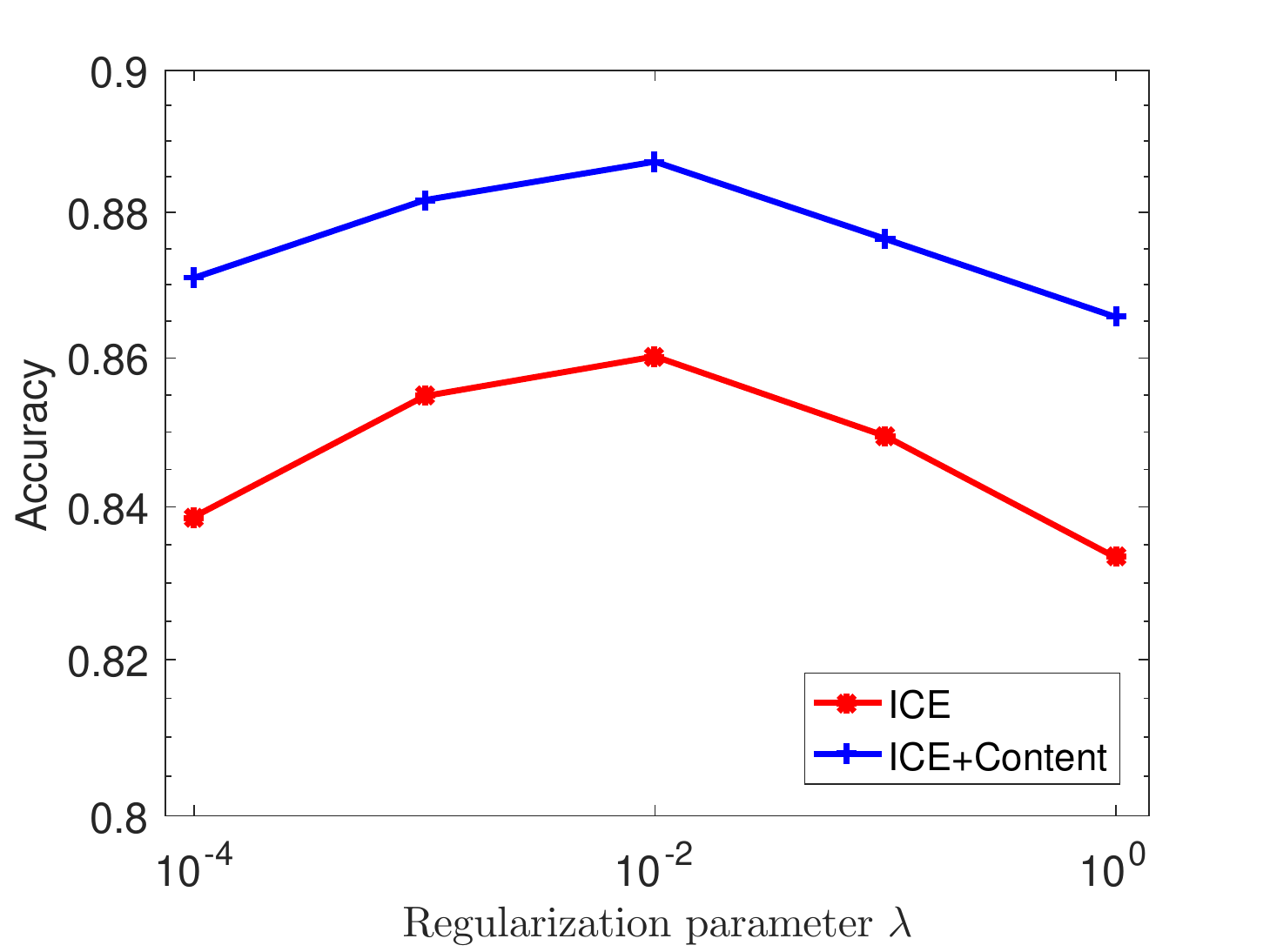}
\label{fig:regularization}
\end{minipage}
}
\caption{Performance of ICE with varying parameters evaluated by $Accuracy$.}
\label{fig:parameter}
\end{figure*}

\subsection{Performance Comparison}

To investigate the performance of ICE and compared methods, we conduct experiments on the Weibo dataset, and report the $Accuracy$, $Precision$, $Precision$, $F1$-$score$, and $Precision$-$Recall$ curves of these methods.

Table \ref{tab:performance} illustrates the performance comparison with the dimensionality $d=8$ on the Sina Weibo dataset evaluated by $Accuracy$, $Precision$, $Precision$ and $F1$-$score$. Using part of the features, NewsCP-Content and NewsCP-Social have the lowest $Accuracy$ and $F1$-$score$ among all the methods. Meanwhile, we can see that the performance of NewsCP-Social is better than that of NewsCP-Content. This may indicate that social features are more important than content features for evaluating the credibility. Involving both kinds of features, NewsCP achieves great improvement and has a satisfactory performance. Then, EP further improves the performance compared to NewsCP and achieves an $Accuracy$ of more than $80\%$. Incorporating content information of NewsCP-Content, EP+Content achieves a little improvement and becomes the best one among all the compared methods. We can clearly observe that, our proposed ICE model outperforms the compared methods. Incorporating content features, ICE+Content achieves the best performance among all the methods. Compared to EP+Content, ICE and ICE+Content improve the Accuracy by $3.7\%$ and $6.4\%$, respectively. When the target class is rumor, the F1-score improvements are $3.0\%$ and $5.1\%$, respectively, and when the target class is non-rumor, the improvements are $3.5\%$ and $6.0\%$, respectively. Moreover, among the results of all the methods, the F1-score of rumors is higher than the F1-score of non-rumors. This may mean that it is more difficult to distinguish non-rumors from rumors than to distinguish rumors from non-rumors.

We also illustrate the $ Precision $-$ Recall $ curves of the different methods in Figure \ref{fig:PR}. The $ Precision $-$ Recall $ curve for rumors in Figure \ref{fig:PR1} shows that ICE+Content outperforms other methods. The $ Precision $ of ICE+Content stays $100\%$ until $ Recall $ is more than $50\%$. ICE is better than other compared methods in most cases, except at around $recall = 45\%$. The $ Precision $-$ Recall $ curve for non-rumors in Figure \ref{fig:PR2} shows that the performance of ICE and ICE+Content is clearly better than that of other methods. The $ Precision $ of ICE stays $100\%$ until its $ Recall $ is more than $40\%$. The $ Precision $ of ICE+Content stays $100\%$ until its $ Recall $ is more than $70\%$. In our experiments, both experimental results in Table \ref{tab:performance} and $ Precision $-$ Recall $ curves in Figure \ref{fig:PR} clearly show that our proposed ICE model achieves satisfactory performances and outperforms the other state-of-the-art methods.

\subsection{Impact of Parameters}

To investigate the impact of parameters on the performance of ICE, we illustrate the $Accuracy$ performance of ICE with varying parameters in Figure \ref{fig:parameter}. Based on the figure, we can select the best parameters for ICE.

In Figure \ref{fig:dimensionality}, we illustrate the performance of ICE and ICE+Content with varying dimensionality $d$, where the regularization parameter is set to be $\lambda =0.01$. The performance of ICE increases rapidly from $d=3$ and then becomes stable since $d=6$. ICE achieves the best performance at $d=8$ and then decreases slightly with the increasing dimensionality. Meanwhile, the performance of ICE+Content has the similar trend. However, the $Accuracy$ curve of ICE+Content is more stable than that of ICE, because the performance of modeling content information is not influenced by the dimensionality. From the observation of their curves, we select the best dimensionality of ICE as $d=8$. Moreover, their curves show that ICE is not very sensitive to the dimensionality in a large range, and ICE still outperforms the compared methods even not with the best dimensionality.

\begin{table*}[tb]
  \centering
  \caption{Performance comparison under different event topics and event popularity evaluated by $Accuracy$.}
    \begin{tabular}{c|ccccc|cccc}
    \toprule
    \multirow{2}[0]{*}{Methods} & \multicolumn{5}{c|}{Topics}           & \multicolumn{4}{c}{Popularity} \\
          & Society & Life  & Policy & Politics & Entertainment & $0$-$100$ & $100$-$300$ & $300$-$1000$ & $1000$+ \\
    \midrule
    NewsCP-Content & 0.608  & 0.524  & 0.667  & 0.541  & 0.676  & 0.662  & 0.556  & 0.568  & 0.611  \\
    NewsCP-Social & 0.649  & 0.571  & 0.667  & 0.568  & 0.595  & 0.647  & 0.556  & 0.595  & 0.639  \\
    NewsCP & 0.797  & 0.619  & \textbf{0.933} & 0.703  & 0.730  & 0.794  & 0.689  & 0.757  & 0.778  \\
    EP    & 0.811  & 0.857  & 0.800  & 0.811  & 0.811  & 0.824  & 0.822  & 0.811  & 0.778  \\
    EP+Content & 0.811  & 0.857  & 0.867  & 0.838  & \textbf{0.838} & 0.853  & 0.844  & 0.811  & 0.778  \\
    \textbf{ICE}   & 0.838  & 0.905  & \textbf{0.933} & 0.865  & \textbf{0.838} & 0.868  & 0.867  & \textbf{0.865} & \textbf{0.861} \\
    \textbf{ICE+Content} & \textbf{0.851} & \textbf{0.952} & \textbf{0.933} & \textbf{0.946} & \textbf{0.838} & \textbf{0.882} & \textbf{0.933} & \textbf{0.865} & \textbf{0.861} \\
    \bottomrule
    \end{tabular}%
  \label{tab:situations}%
\end{table*}%

The $Accuracy$ curves of ICE and ICE+Content with varying regularization parameter $\lambda$ are shown in Figure \ref{fig:regularization}. The performance of ICE grows slowly from $\lambda=0.0001$ and achieves the highest $Accuracy$ at $\lambda=0.01$. It is obvious that the best parameter for ICE is $\lambda=0.01$. We can also clearly observe that ICE stays stable in the range of $\lambda$ from $0.0001$ to $1$. Moreover, recalling results in Table \ref{tab:performance}, even not with best parameters, the performances of ICE are still better than those of the compared methods.

\subsection{Performance Under Different Situations}

\begin{figure}[tb]
\centering
\subfigure[Distribution of different topics.]{
\begin{minipage}[b]{0.22\textwidth}
\includegraphics[width=1\textwidth]{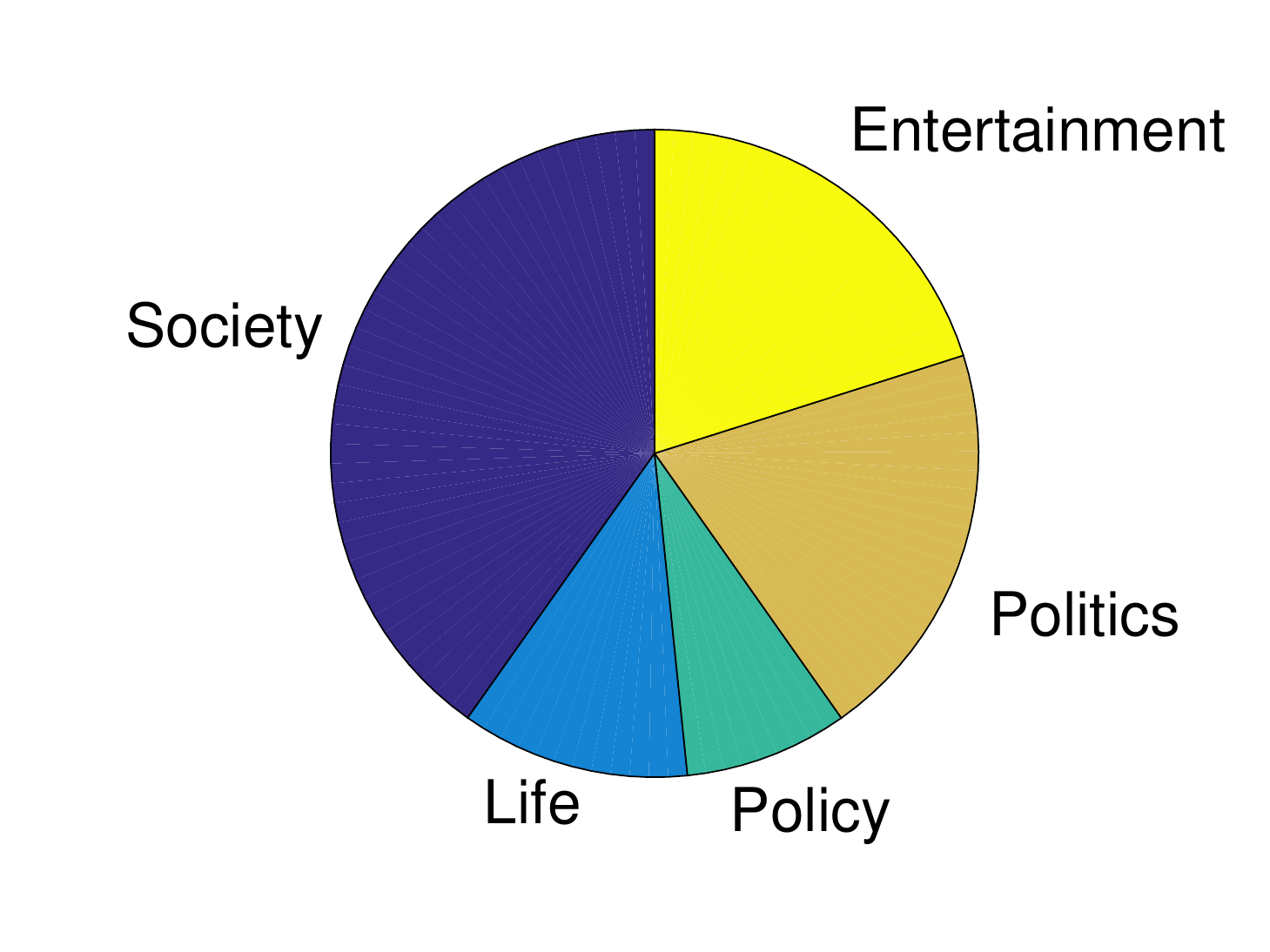}
\label{fig:FB1}
\end{minipage}
}
\subfigure[Distribution of different levels of popularity.]{
\begin{minipage}[b]{0.22\textwidth}
\includegraphics[width=1\textwidth]{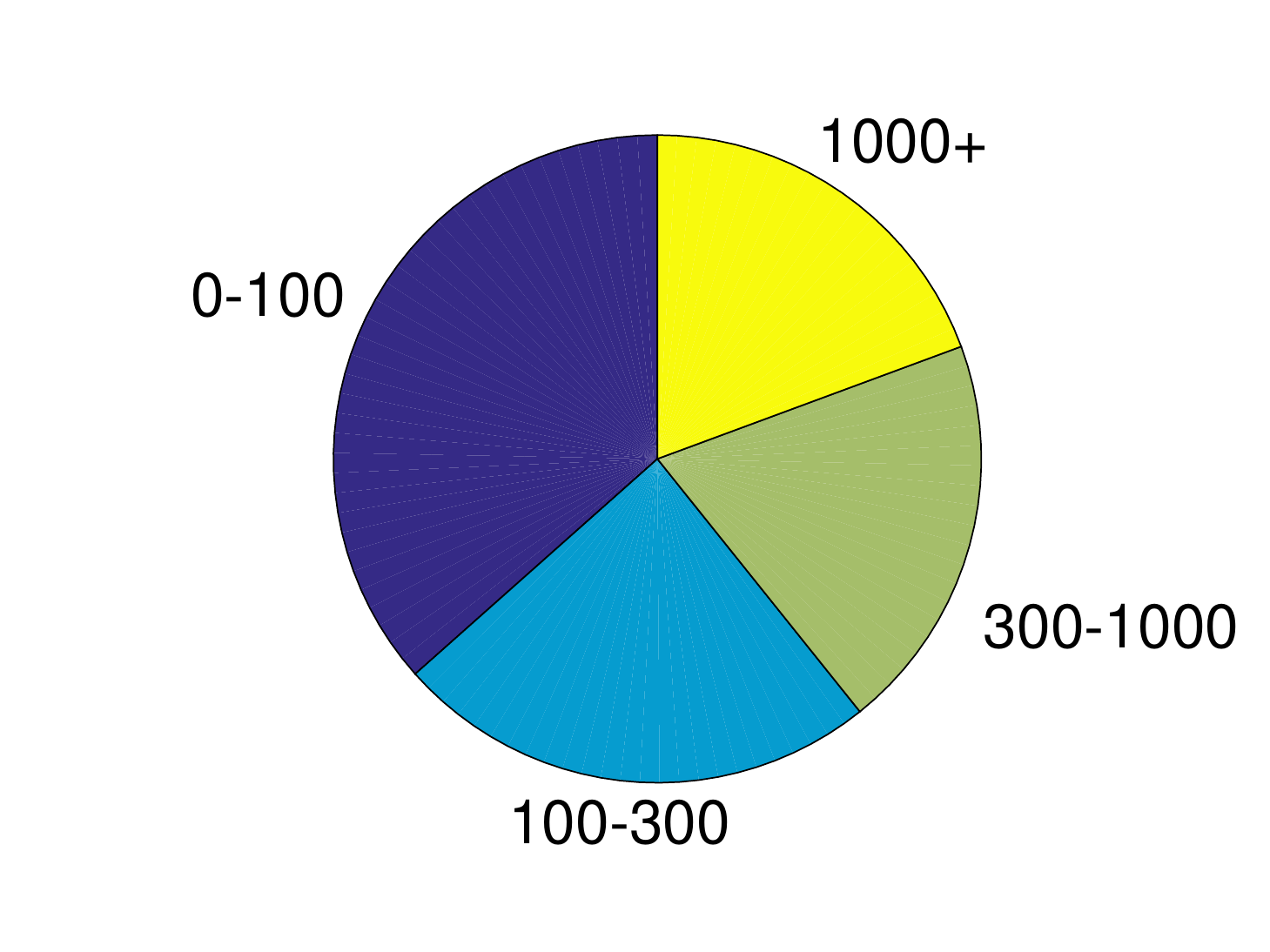}
\label{fig:FB2}
\end{minipage}
}
\caption{Distribution of different situations in the Weibo datatset.}
\label{fig:FB}
\end{figure}

We have shown that the proposed ICE model can outperform the state-of-the-art methods. Here, we are going to investigate if ICE can perform better than the compared methods in some specific situations. We partition our dataset according to topics and popularity, and the results evaluated by $Accuracy$ under different situations are shown in Table \ref{tab:situations}. The distribution of different situations is shown in Figure \ref{fig:FB}.

According to topics of events, we first partition the dataset into five categories: society, life, policy, politics and entertainment, as shown in Figure \ref{fig:FB1}. Topic ``society" talks about all kinds of stuff happening around us, topic ``life" contains life skills such as health tips, topic ``policy" denotes news about newly released policies, topic ``politics" talks about politics, government and military, and topic ``entertainment" means news about movies, music and sports. Table \ref{tab:situations} illustrates the $Accuracy$ of different methods on the five topics. The results show that our proposed ICE model outperforms the compared methods on all the five topics, and ICE+Content achieves the best performance in all the situations. Meanwhile, NewsCP performs well on the topic of ``policy", and EP has a good performance on the topic of ``entertainment". Moreover, in average, these methods achieve slightly poor performances on topics of ``society" and ``entertainment" compared to other topics. This may indicate that news about ``society" and ``entertainment" has more noise and such rumors are difficult to be well identified.

Then, we partition the dataset according to the popularity of events. The popularity of an event is computed as the amount of its microblogs, including postings and repostings. As shown in Figure \ref{fig:FB2}, the whole dataset is partitioned into four categories: $0$-$100$, $100$-$300$, $300$-$1000$, and $1000$+. From Table \ref{tab:situations}, we can clearly observe that with all kinds of popularity, ICE outperforms the compared methods, and ICE+Content achieves the best performance. Moreover, we can observe that the larger the popularity, the lower the accuracy in most cases. It may be because that the majority of microblogs contain noise and do not contribute to the evaluation very much. Among the massive microblogs on social media, several significant microblogs are easily hidden by a large amount of noise. Thus, in future work, we need to find a method to select significant microblogs, instead of average calculation.

\begin{figure}[tb]
\centering
\includegraphics[width=1\linewidth]{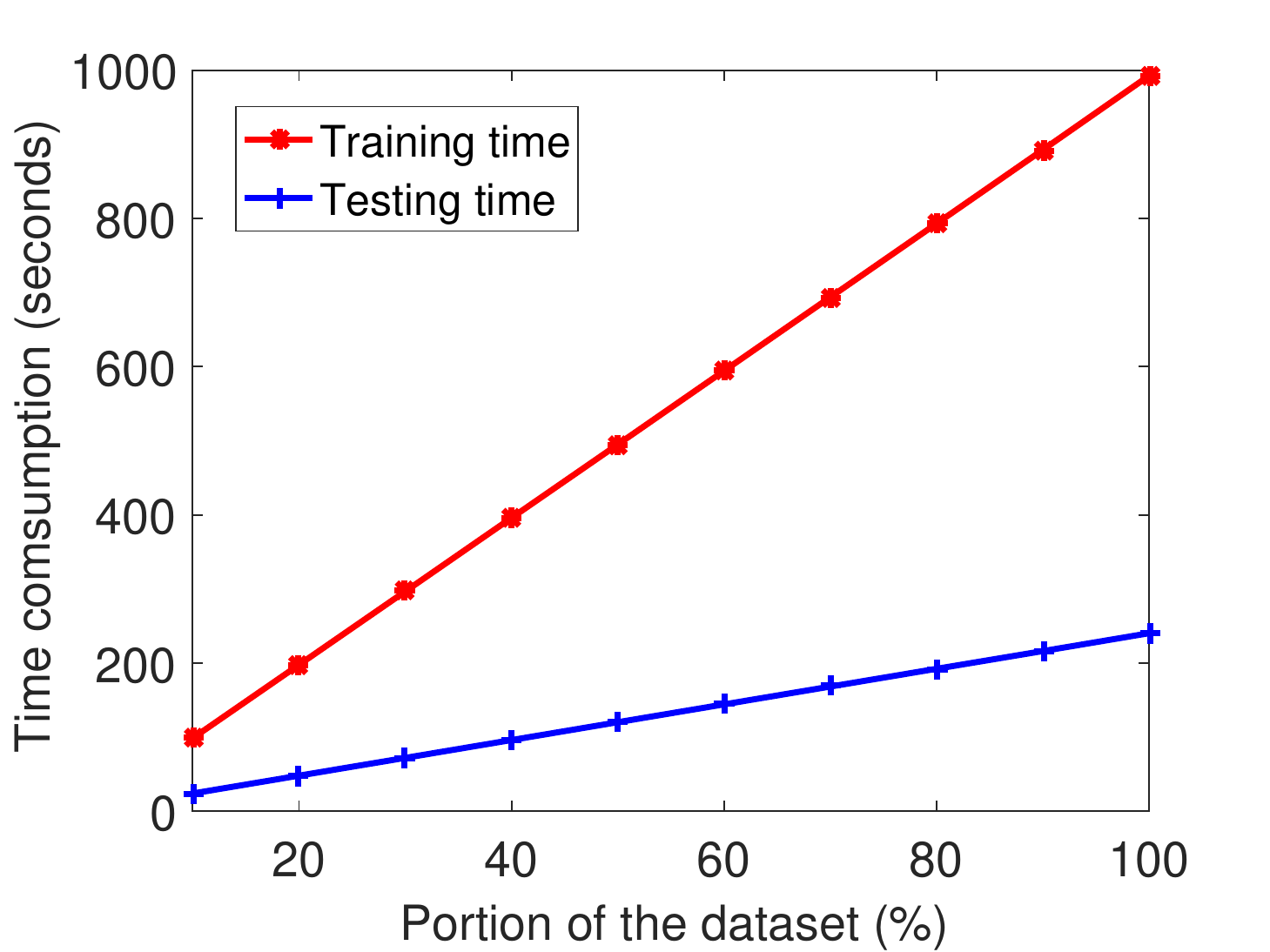}
\caption{Training and testing time consumption of ICE with varying portion of the whole Weibo dataset.}
\label{fig:runtime}
\end{figure}

\subsection{Scalability Analysis}

Besides the analysis of the effectiveness of ICE, we investigate the scalability of the ICE model with varying portions of the Weibo dataset. The model is implemented with Python\footnote{https://www.python.org/.} and Theano\footnote{http://deeplearning.net/software/theano/.}. The code is run on a computer with 4 Core 2.5 GHz CPU and 16 GB RAM, and the GPU model is NVIDIA Tesla K20Xm. On the Weibo dataset, we measure the corresponding time cost of one iteration in both training and testing process. Figure \ref{fig:runtime} shows the time consumption with varying portions of the whole dataset. We can observe that both training and testing time consumption of ICE are linear with respect to the size of dataset. This shows the scalability of ICE. Our proposed model not only can achieve the state-of-the-art performance on credibility evaluation, but also can run effectively on large-scale data.

\begin{figure*}[tb]
\centering
\includegraphics[width=1\linewidth]{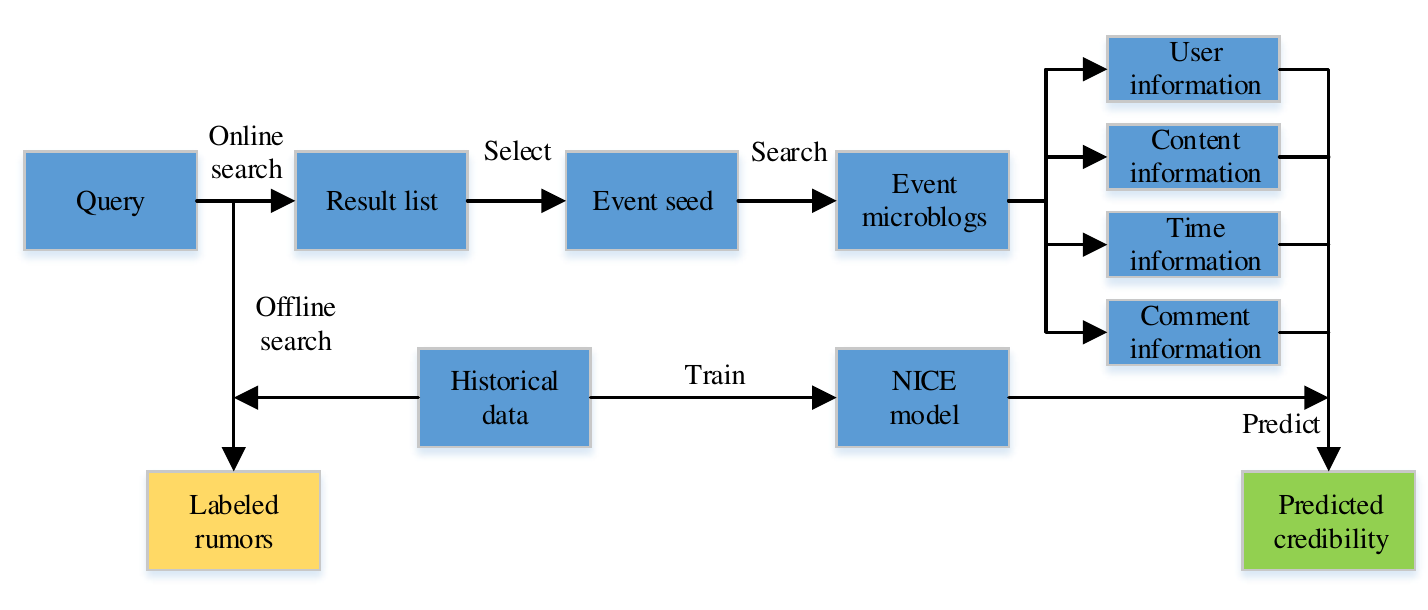}
\caption{Overview of the Network Information Credibility Evaluation (NICE) system.}
\label{fig:overview}
\end{figure*}

\section{System}

As introduced and discussed above, we have achieved a information credibility evaluation model ICE with the state-of-the-art performance. Rather than only using the ICE model in academic datasets and research, it is vital to construct a real-time information credibility evaluation system on social media and make our proposed model applied in real applications. Thus, based on our model and the Sina Weibo dataset, we built a Network Information Credibility Evaluation (NICE) system \cite{wu2016information}. NICE is a webpage-based system that can automatically crawl online information from Sina Weibo and evaluate the credibility of online information that users enquire.

Figure \ref{fig:overview} illustrates the flow chart of the NICE system. Using the system, a user can input a query to retrieve related information. If a user's query matches rumors in the Weibo dataset, users can identify the rumor immediately. Otherwise, NICE will crawl real-time information from social media, i.e., Sina Weibo. Then, the user can select one microblog to evaluate the corresponding credibility based on our model. Based on the selected microblog, the system will crawl all the related microblogs from Weibo and collect related information including content information, temporal information, comment information, and corresponding user profiles. Based on this information and the trained model, NICE can evaluate the credibility of all the related information and provide a predicted score of the event. With our proposed ICE model, the NICE system can achieve great performance in information credibility evaluation. It can be applied effectively and stably in online information management on social media.

\section{Conclusions and Future Work}

In this work, to evaluate information credibility on social media, a novel method, i.e., ICE, has been proposed. ICE aims to learn dynamic representations for the microblogs that describe events spreading on social media. The learning is based on the user credibility, behavior types, temporal properties, and comment attitudes. The aggregation of these key factors makes the dynamic and joint representations of microblogs, and the aggregation of representations of all the microblogs during information spreading can generate the credibility representation of events on social media. Experiments conducted on a real dataset crawled from Sina Weibo show that ICE outperforms the state-of-the-art methods.

In the future, we can further investigate the following directions. First, in ICE, the content information has not been considered when learning credibility representations. We plan to analyze the event content and extract its main elements to predict the happening probability of the event based on a large news database. Second, information about an event on other platforms, e.g., news websites and forums, can be incorporated in our model. Third, for the aggregation of microblogs of an event, we use average computation in ICE, which is clearly not the best solution. We need to find a method to select the significant microblogs.



\balance
\bibliographystyle{abbrv}
\bibliography{IEEE}

\end{document}